\documentclass[prb,groupedaddress,superscriptaddress,twocolumn,footinbib]{revtex4-1}

\usepackage{graphicx}
\usepackage{dcolumn}
\usepackage{bm}
\usepackage[T1]{fontenc}
\usepackage[french]{babel}
\usepackage{epstopdf}
\usepackage{amssymb}
\usepackage{color}
\usepackage{amsmath}
\usepackage{subfigure}
\usepackage{natbib}

\begin{document}

\preprint{}

\title{Graphene bubbles on a substrate: Universal shape and van der Waals pressure}
\author{E. Khestanova}
\affiliation{School of Physics and Astronomy, University of Manchester, Oxford Road, Manchester M13 9PL, UK}
\author{F. Guinea$^*$}
\affiliation{School of Physics and Astronomy, University of Manchester, Oxford Road, Manchester M13 9PL, UK}
\affiliation{IMDEA Nanociencia, Faraday, 9, Cantoblanco, 28049, Madrid, Spain}
\author{L. Fumagalli}
\affiliation{School of Physics and Astronomy, University of Manchester, Oxford Road, Manchester M13 9PL, UK}
\author{A. K. Geim}
\affiliation{School of Physics and Astronomy, University of Manchester, Oxford Road, Manchester M13 9PL, UK}
\author{I. V. Grigorieva$^*$}
\affiliation{School of Physics and Astronomy, University of Manchester, Oxford Road, Manchester M13 9PL, UK}

\begin{abstract}
Trapped substances between a 2D crystal, such as graphene, and an atomically flat substrate, for example, hexagonal boron nitride, give rise to the formation of bubbles. We show that the size, shape and internal pressure inside these bubbles are determined by the competition between van der Waals attraction of a 2D crystal to the substrate and the elastic energy needed to deform the atomically thin layer. This presents opportunities to use bubbles to study the elasticity of 2D materials as well as the conditions of confinement, yet none of these have been explored so far, either theoretically or experimentally. We have created a variety of bubbles formed by monolayers of graphene, hBN and MoS$_2$ mechanically exfoliated onto hBN, graphite and MoS$_2$ substrates. Their shapes, analyzed using atomic force microscopy, are found to exhibit universal scaling with well-defined aspect ratios, in agreement with theoretical analysis based on general properties of membranes. We also measured the pressure induced by the confinement, which increased with decreasing bubble's size and reached tens on MPa inside submicron bubbles. This agrees with our theory estimates and suggests that for bubbles with radii $\lesssim10$ nm hydrostatic pressures can reach close to 1 GPa, which may modify the properties of a trapped material.
\end{abstract}

\maketitle

Van der Waals heterostructures$^1$ - stacks of atomically thin layers of different materials assembled layer by layer - are making possible the design of new devices with tailored properties. An essential feature of such heterostructures is atomically clean interfaces that form due to strong adhesion between the constituent layers$^2$. Even though contamination (adsorbed water, hydrocarbons) is inevitably present on individual layers before assembly, the van der Waals (vdW) forces that attract adjacent two-dimensional (2D) crystals squeeze out trapped contaminants, usually pushing them into submicron-size 'bubbles' and leaving large interfacial areas atomically sharp and free of contamination$^2$.

So far such bubbles have been used simply as signatures of good adhesion between constituents of vdW heterostructures and as indicators that the interfacial areas between the bubbles are perfectly clean$^3$. Now we show that the bubbles can be employed as a tool to study the elastic properties of the 2D crystals involved and, also, to evaluate the conditions that nanoscale confinement exerts on the enclosed material (e.g., hydrostatic pressure).
This information is important in many situations where confinement can modify materials properties, with water inside graphene nanocapillaries$^{4-6}$, nanocrystals or biological molecules confined in graphene liquid cells$^{7-9}$, room-temperature ice in a 2D nanochannel$^{10,11}$ and a 'hydrothermal anvil' made of graphene on diamond$^{12}$ being a few examples. Furthermore, highly strained graphene nanobubbles have been shown to possess enormous pseudo-magnetic fields$^{13}$, greater than 300 T. The detailed knowledge of strain for commonly occurring bubbles should facilitate studies of the electronic properties of graphene under conditions inaccessible in high-field magnet laboratories$^{14}$. 

Here we study bubbles formed between a 2D crystal (monolayer graphene, monolayer hexagonal boron nitride (hBN) or monolayer MoS$_2$) and an atomically smooth flat substrate (hBN, graphite, MoS$_2$). By analyzing shapes and dimensions of the bubbles and comparing them with the corresponding predictions of the elasticity theory we find that the bubbles for all three materials are fully described by the combination of a 2D crystal's elastic properties and its vdW attraction to a substrate. We find excellent agreement between experiment and theory, both for smoothly deformed bubbles and for bubbles with shape and dimensions modified by a residual strain. Furthermore, using indentation of bubbles with an AFM tip, we extracted the hydrostatic (vdW) pressure inside them and Young's moduli for graphene and MoS$_2$ membranes.

\section{Experiment}
Samples for this study were made by mechanical exfoliation of graphene, hBN and MoS$_2$ monolayers onto hBN, graphite, and MoS$_2$ substrates. This resulted in spontaneous formation of a large number of bubbles filled with hydrocarbons$^{2}$, with typical separations from $\sim{0.5}$ to tens of microns. To ensure that the prepared heterostructures reach equilibrium conditions, they were annealed at 150$^{\circ}$C for 20-30 minutes. After that the dimensions and topography of many bubbles (up to 100 for each heterostructure) were analyzed using atomic-force microscopy (AFM). 

\begin{figure}[h]
	\includegraphics[width=0.5\textwidth]{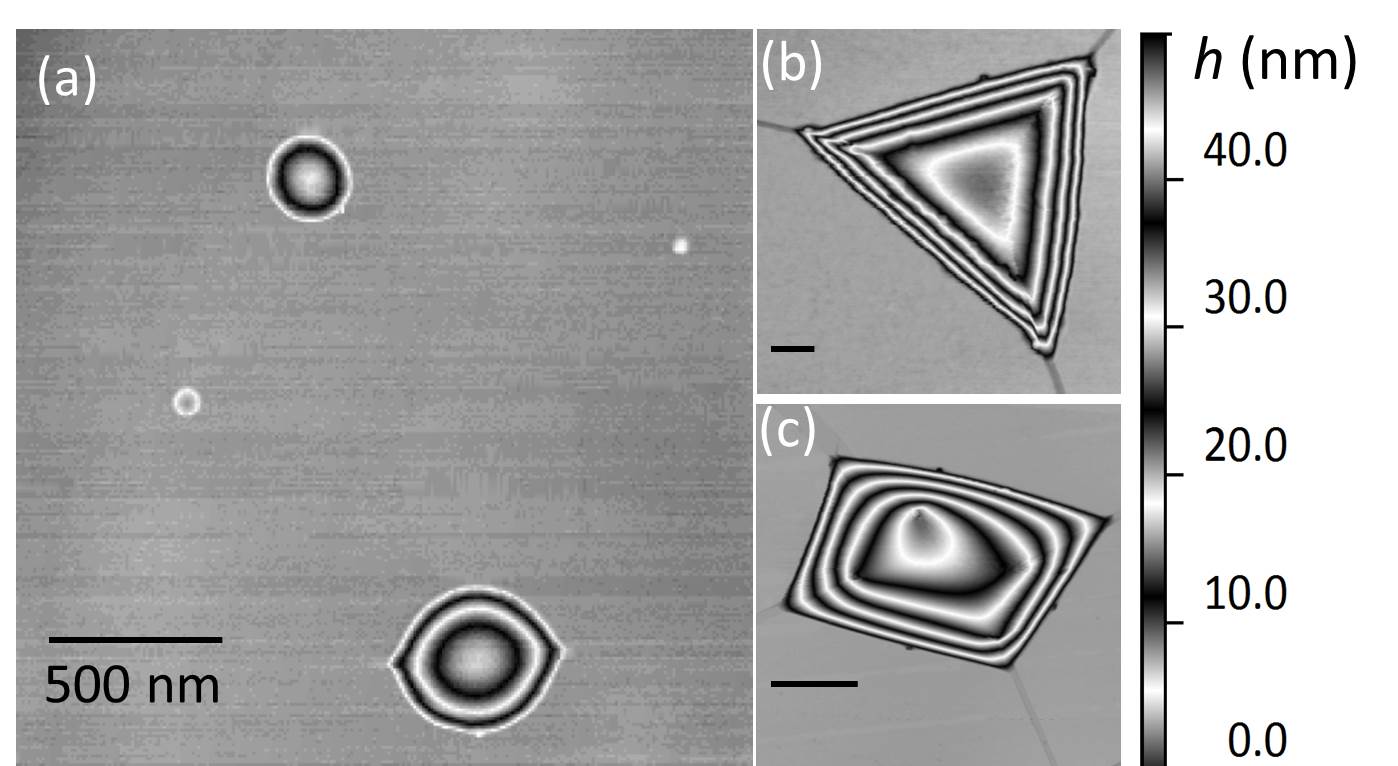}
	\caption{\label{fig:bubbles_images} \textbf{Graphene bubbles}. (a)-(c) AFM images of graphene bubbles of different shapes. Scale bars in (b) and (c) correspond to 100 and 500 nm, respectively. The vertical scale on the right indicates the height of the bubbles. }
\end{figure}

Fig. \ref{fig:bubbles_images} shows typical examples of bubbles formed by monolayer graphene on bulk hBN. The majority of the bubbles were smaller than 500 nm in radius, $R$, and had a round or nearly round base (Fig. \ref{fig:bubbles_images}a). Larger bubbles typically exhibited pyramidal shapes, with either triangular (Fig. \ref{fig:bubbles_images}b) or trapezoidal (Fig. \ref{fig:bubbles_images}c) bases.

Bubbles formed by monolayer hBN on bulk hBN were also either round or approximately triangular in shape but  smaller in size compared to graphene (<100 nm for round and <500 nm for triangular bases). Bubbles formed by MoS$_{2}$ monolayers were mostly round, similar to those shown in Fig. \ref{fig:bubbles_images}a for graphene, but exhibited a broader size distribution, with $30 <R <1000$ nm. We measured the cross-sectional profiles of the observed bubbles and analyzed their maximum height, $h_{max}$, and the aspect ratio of $h_{max}$ to the radius, $R$, or to the length of the side, $L$, as appropriate.

\begin{figure}[h]
 	\includegraphics[width=0.5\textwidth]{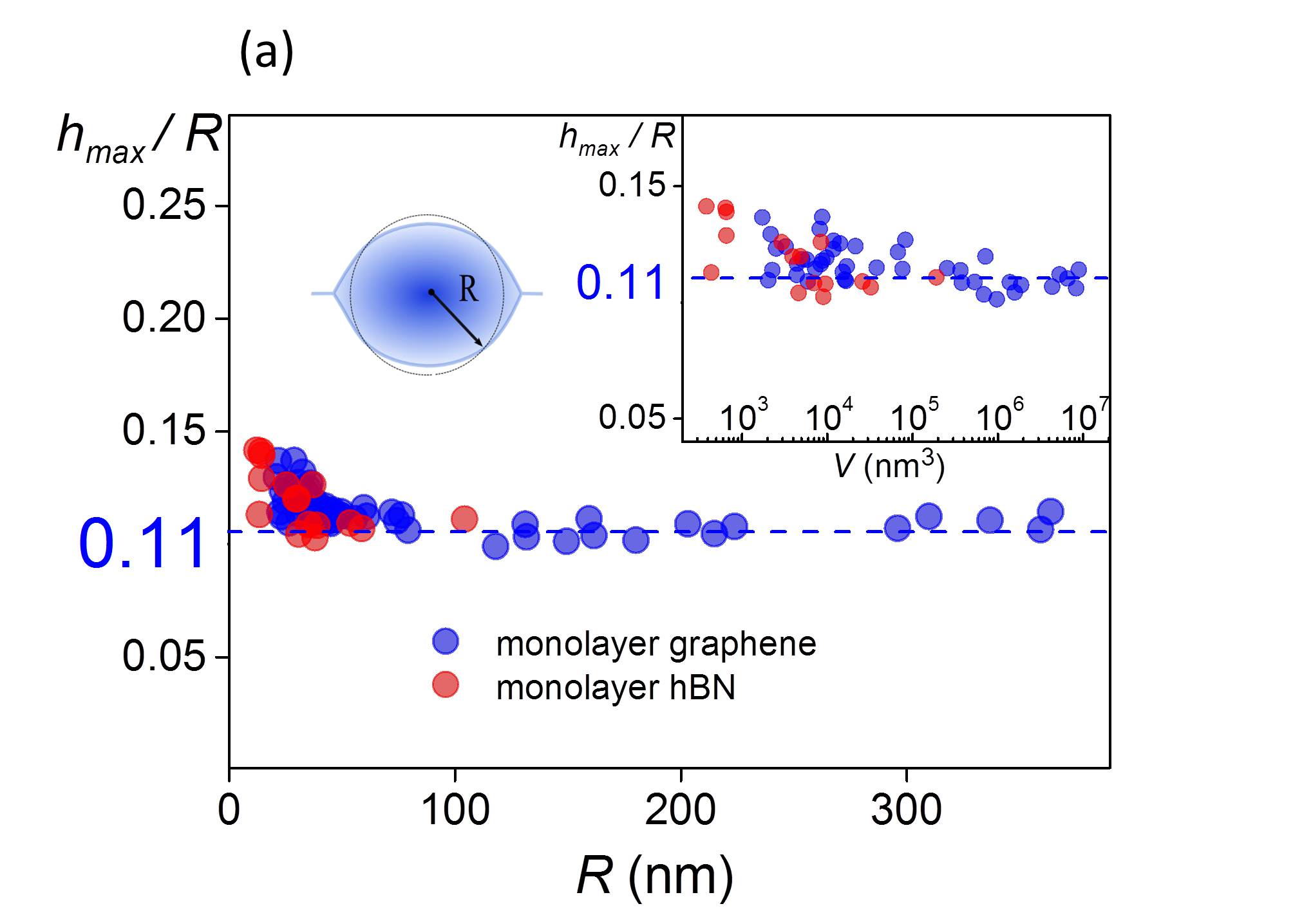}
 	\includegraphics[width=0.5\textwidth]{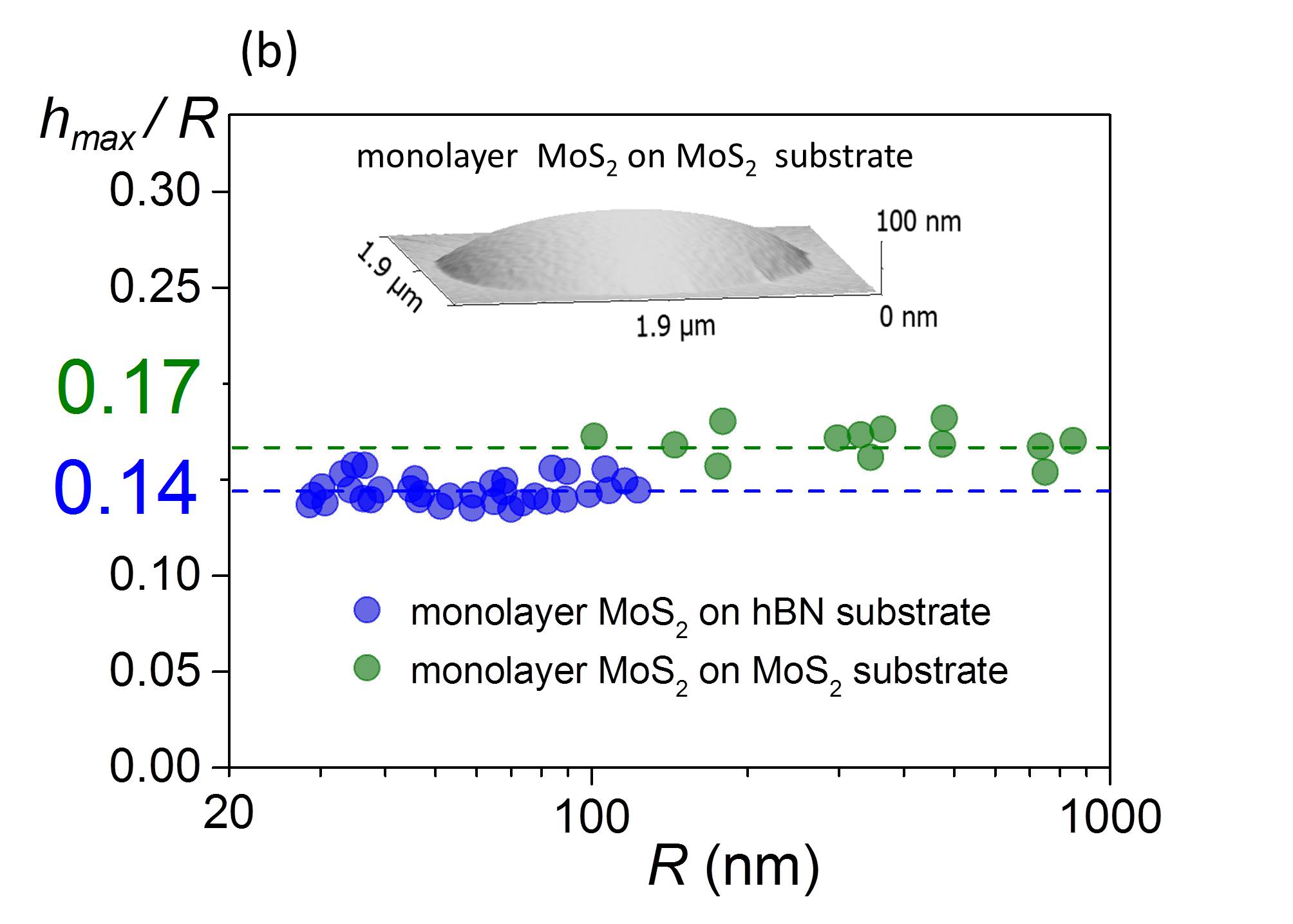}
 	\caption{\label{fig:bubbles_scaling_graphene_MoS2_round} \textbf{Universal shape of round-type bubbles}. (a) Measured aspect ratios as a function of the base radius for graphene (blue symbols) and monolayer hBN (red symbols). Dashed line shows the mean value. Top left inset: sketch of a nearly round bubble and its effective radius $R$ determined as $R=\sqrt{A/\pi}$, where $A$ is the measured area of the base of the bubble. Right inset: aspect ratio of the bubbles as a function of their volume. (b) Aspect ratio of MoS$_2$ bubbles on hBN and MoS$_2$ substrates. Dashed lines show the mean values of $h_{max}/R=0.14$ and $0.17$, respectively. The logarithmic scale is used to accommodate the large range of $R$. Inset: AFM image of a typical MoS$_2$ bubble.}
 \end{figure}

The results for round-type graphene bubbles are shown in Fig. \ref{fig:bubbles_scaling_graphene_MoS2_round}a. The aspect ratio, $h_{max}/R$, is remarkably universal, that is, independent of the bubbles' radius, $R$, or volume, $V$: $h_{max}/R\approx0.11$, within $\sim10\%$. Moreover, if we discount the smallest bubbles with $R<50$ nm, the accuracy reaches 4\% for sizes varying by an order of magnitude. 
Very similar behavior was found for monolayer hBN, with $h_{max}/R\approx0.11$ for bubbles larger than 50 nm and a somewhat increasing $h_{max}/R$ for $R<50$ nm - see Fig. \ref{fig:bubbles_scaling_graphene_MoS2_round}a. Only a few sufficiently large bubbles were found in this case, limiting our analysis.

Aspect ratios for round bubbles formed by MoS$_2$ monolayers are shown in Fig. \ref{fig:bubbles_scaling_graphene_MoS2_round}b. For comparison we analyzed the MoS$_2$ bubbles formed on two different substrates, MoS$_2$ and hBN. Again, for the same 2D crystal--substrate combination we find a constant $h_{max}/R$ but its value depends on the substrate and is notably larger compared to graphene and hBN monolayers. This can be attributed to different elastic properties of monolayer MoS$_2$ compared to one-atom-thick crystals (graphene, monolayer hBN). Furthermore, the different $h_{max}/R$ found for different substrates point at the importance of vdW adhesion, as discussed below.

\begin{figure}[h]
	\includegraphics[width=0.5\textwidth]{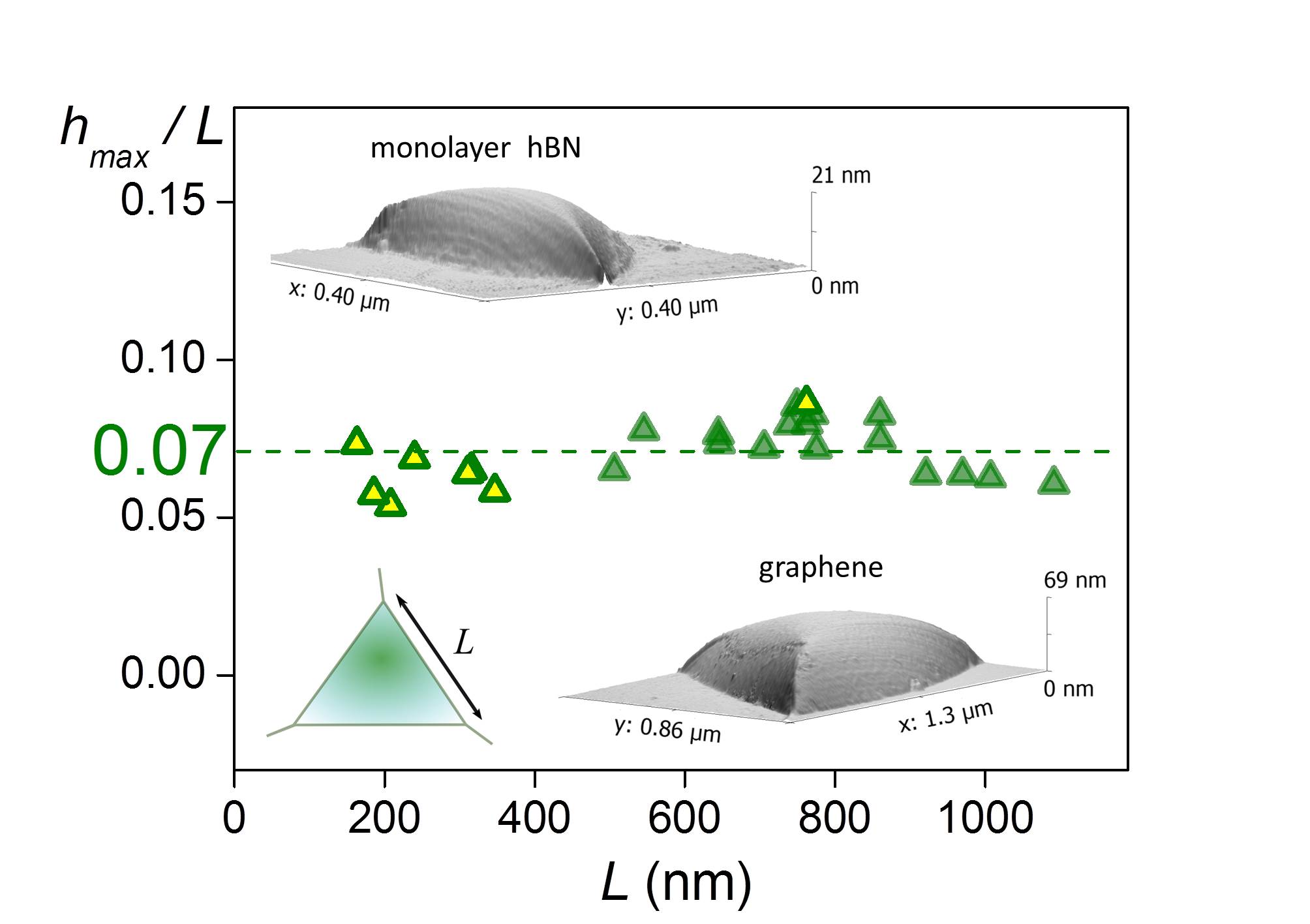}
	\caption{\label{fig:bubbles_scaling_graphene_trapez_smooth} \textbf{Aspect ratio of smooth triangular bubbles}. Symbols show  the measured aspect ratios of graphene and hBN bubbles (closed and open symbols, respectively), both on hBN substrates, as a function of $L$. The dashed line shows the mean aspect ratio, $h_{max}/L=0.07$. Bottom left inset: Sketch of a triangular bubble. Its side length $L$ was experimentally determined as $L=\sqrt{4A/\sqrt{3}}$, where $A$ is the measured area of the base of a bubble. The other two insets show typical AFM images of smoothly deformed triangular bubbles.}
\end{figure}

A constant aspect ratio was also found for graphene bubbles with triangular bases such as those shown in Fig. \ref{fig:bubbles_images}b and Fig. \ref{fig:bubbles_scaling_graphene_trapez_smooth}. In this case it is intuitive to use the length of the side, $L$, to characterize their sizes. Similar to the round bubbles in Fig. \ref{fig:bubbles_scaling_graphene_MoS2_round}a, these bubbles usually had smooth round tops but were larger in size (typical $L$ between 500 and 1000 nm) and exhibited the aspect ratio $h_{max}/L=0.07\pm0.01$ - see Fig. \ref{fig:bubbles_scaling_graphene_trapez_smooth}. We note that, although this  value appears to be lower than that for the round bubbles, as if the triangular bubbles were somewhat thinner, this is simply the effect of using a different measure to characterize the lateral size ($L$ vs $R$). Indeed, redefining the lateral size of triangular-type bubbles as a distance $L^*$ from their centers to corners, we find the same ratio $h_{max}/L^*$ as for round bubbles, within our experimental accuracy. As discussed below, the shapes and dimensions of all smoothly deformed bubbles (round or triangular) are expected to follow the same scaling.

\begin{figure}[h]
	\includegraphics[width=0.5\textwidth]{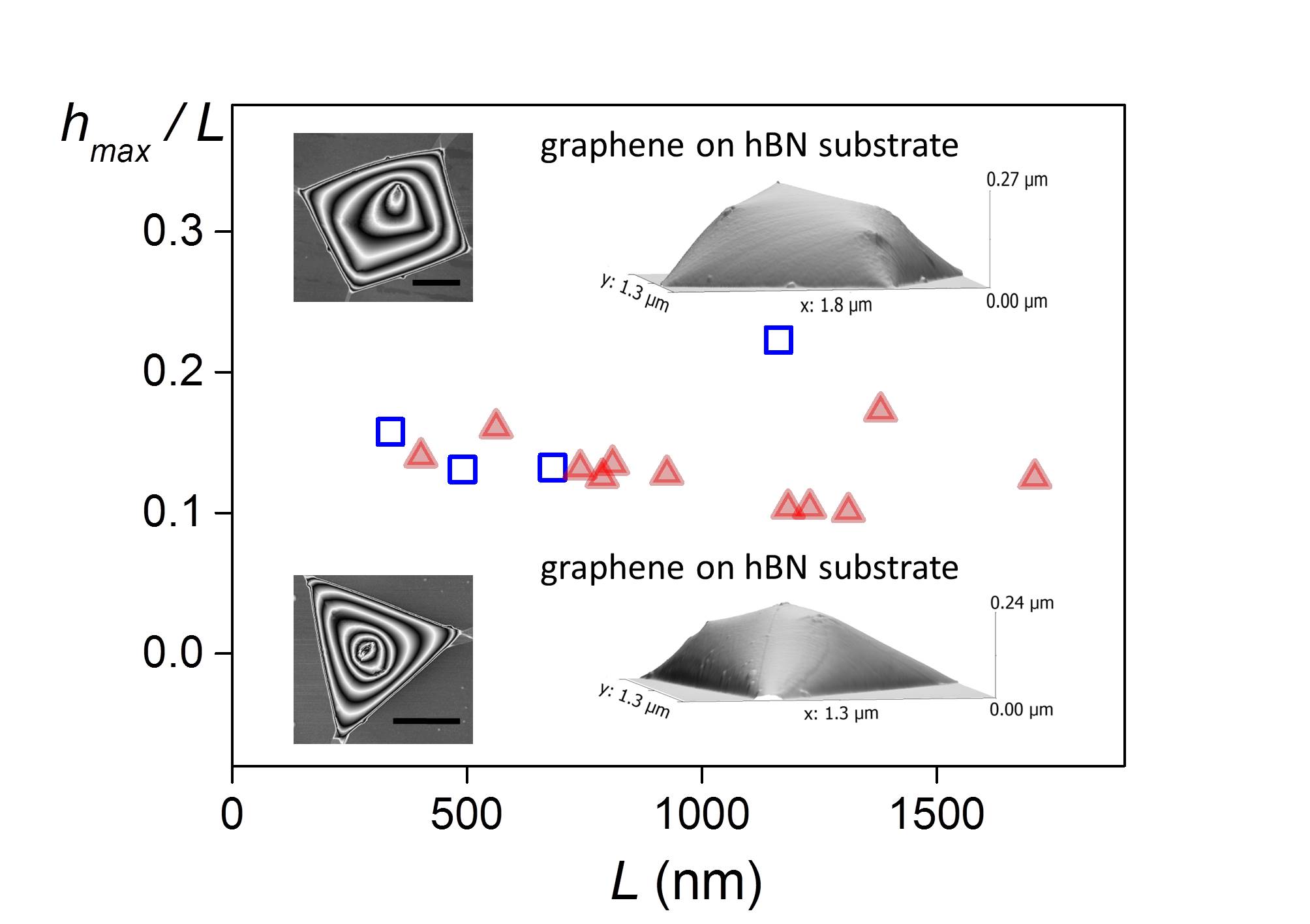}
	\caption{\label{fig:bubbles_scaling_graphene_trapez}\textbf{Aspect ratio of pyramidal graphene bubbles with sharp features.} Symbols show the measured aspect ratios of triangular (red symbols) and trapezoidal (blue) bubbles, as a function of their side length, $L$, determined as $L=\sqrt{4A/\sqrt{3}}$ for triangular bubbles and as $L=\sqrt{A}$ for trapezoidal ones. Here $A$ is the measured area of the base of a bubble. Insets show typical AFM images of such bubbles: left and right are top and 3D views, respectively. Scale bars, 500 nm.}
\end{figure}

The only class of bubbles that showed strong deviations from the universal scaling behavior were pyramidal-type bubbles with sharp features. They exhibited sharp ridges that often extended nearly to the full height of the bubbles. Two examples are shown as insets in Fig. \ref{fig:bubbles_scaling_graphene_trapez}. The aspect ratio, $h_{max}/L$, for such bubbles showed relatively large variations (by a factor of 2), with most values being higher than those for smoothly deformed bubbles - c.f. Fig. \ref{fig:bubbles_scaling_graphene_trapez_smooth} and Fig. \ref{fig:bubbles_scaling_graphene_trapez}.

To summarize, all bubbles - formed by graphene, hBN and MoS$_2$ monolayers - exhibited a small set of shapes (mostly, round and triangular) with a universal aspect ratio. Monolayers of graphene and hBN, that have similar elastic properties, showed the same aspect ratio. The aspect ratio for MoS$_2$, that has a lower elastic stiffness$^{5,16}$, was also constant but its value was up to 50\% higher than for graphene and hBN monolayers. The universal behavior for different 2D crystals points to the definitive role played by their elastic properties, as analyzed in the following sections.

\section{Scaling analysis.}
\begin{figure}[h]
	\includegraphics[width=0.30\textwidth]{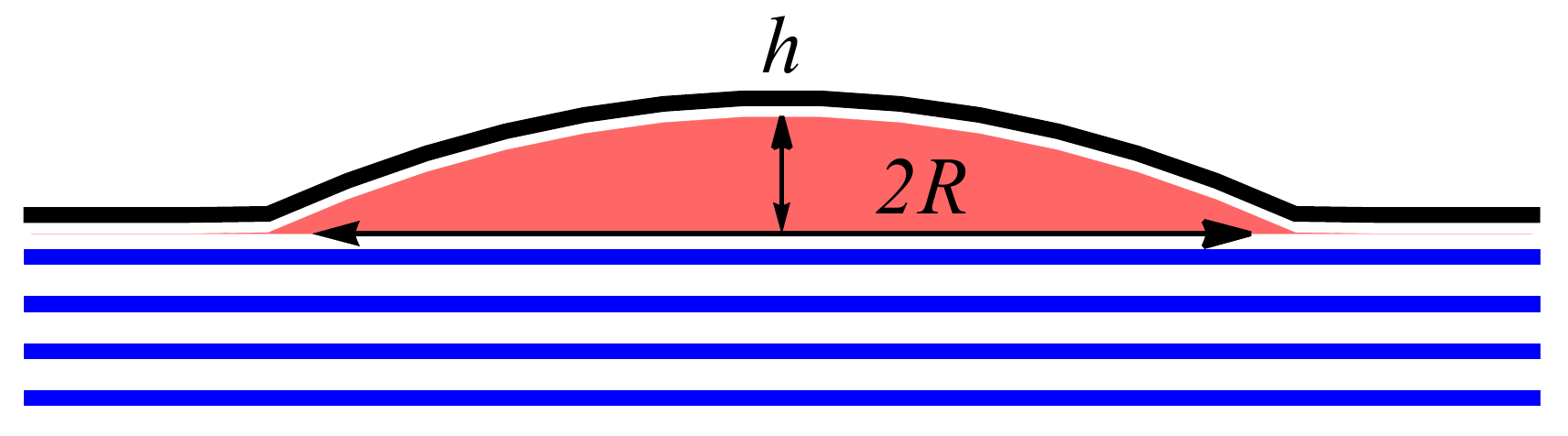}
	\caption{\label{fig:sketch}\textbf{Sketch of the bubble considered in our theoretical analysis.} The bubble is formed by material trapped between a substrate and a 2D layer (graphene).}
\end{figure}

To model the observed bubbles, we consider a material trapped between a flat substrate and a 2D crystal attracted to the substrate by vdW forces (see Fig. \ref{fig:sketch}). For simplicity, we refer to graphene only. Its rigidity is determined by a combination of the in-plane stiffness, and the energy associated with out-of-plane bending. The in-plane stiffness is described by the theory of elasticity$^{17}$, which requires specification of two parameters, Young's modulus,  $Y$, and Poisson's ratio, $\nu$, or, alternatively, Lam\'e coefficients, $\lambda$ and $\mu$. As graphene is an ultimately thin 2D membrane, out-of-plane deformations lead to in-plane stresses, making the system highly anharmonic$^{18}$. The out-of-plane bending is described by the bending rigidity, $\kappa$. Relative contributions of the in-plane stiffness and the bending rigidity to the elastic energy of a 2D membrane are determined by the scale of deformations: Beyond a length scale $\ell_{anh} \sim \sqrt{Y/\kappa}$ the stiffness is dominated by in-plane stresses. For graphene, this scale is $\ell_{anh} \approx 4$ \AA, so that in most situations the bending rigidity can be neglected (however, see further). The equivalent length for MoS$_2$ is somewhat larger, but still smaller than 1 nm.

The vdW energy associated with separating of a graphene layer from the substrate is given by
\begin{align}
E_{vdW} &= \pi \gamma R^2 \nonumber \\
\gamma &= \gamma_{GS} - \gamma_{Gb} - \gamma_{Sb}
\label{EvdW}
\end{align}
where $\gamma_{GS} , \gamma_{Gb}$ and $\gamma_{Sb}$ are the adhesion energies between graphene and the substrate, graphene and the substance inside the bubble, and the substrate and the substance, respectively.

If the bubble is filled with a substance having a finite compressibility, $\beta$, it can be written as
\begin{align}
\beta^{-1} &= V \frac{\partial^2 E_b ( V )}{\partial V^2} = -V \frac{\partial P}{\partial V}
\end{align}
where $E_b ( V )$ is the free energy of the substance inside the bubble of volume $V$, and $P$ is the pressure.

The bubble's height profile is described by
\begin{align}
h ( r ) &= h_{max} \tilde{h} \left( \frac{r}{R} \right)
\end{align}
where $h_{max}$ is the maximum height of the bubble so that $\tilde{h} ( 0 ) = 1 , \tilde{h} ( 1 ) =0$. The in-plane displacements are defined by the function $u_r ( r ) = ( h_{max}^2 / R ) \times \tilde{u}_r  ( R )$. We assume the radial symmetry, so that the azimuthal displacements vanish, i.e., $u_\theta = 0$. Details of calculating the in-plane displacements and the total energy as a function of $h ( r )$ are given in Supplementary Note 1.

Neglecting the bending rigidity, the total energy can be written as
\begin{align}
E_{tot} &= E_{el} + E_{vdW} + E_{b} ( V ) = \nonumber \\ &= c_1 [ \tilde{h} ] Y \frac{h_{max}^4}{R^2} + c_2 [ \tilde{h} ] Y \epsilon h_{max}^2 + \pi \gamma R^2 + E_b ( V )
\label{etot}
\end{align}
where dimensionless coefficients $c_1$ and $c_2$ depend only on the function $\tilde{h}$ describing the height profile, and the volume $V$ is
\begin{align}
V &= c_V [ \tilde{h} ] h_{max} \times R^2.
\end{align}
Below we show that the function $\tilde{h} ( x )$ is generic, i.e., independent of the material parameters $Y , \gamma $ and $E_b ( V )$.

By minimizing equation (\ref{etot}) with respect to $h_{max}$ and $R$ we obtain
\begin{align}
c_1 Y \frac{4 h_{max}^3}{R^2} + 2 c_2 Y \epsilon h_{max} - c_V R^2 P &= 0 \nonumber \\
- c_1 Y \frac{2 h_{max}^4}{R^3} + 2 \pi \gamma R - 2 c_V h_{max} R P &= 0
\label{min}
\end{align}
where we have used $P = - \partial E_b / \partial V$. By eliminating $P$ in equations (\ref{min}), we obtain
\begin{align}
5 c_1 Y \left( \frac{h_{max}}{R} \right)^4 + 2 c_2 Y \epsilon \left( \frac{h_{max}}{R} \right)^2 - \pi \gamma &= 0
\label{hr}
\end{align}
This equation defines the aspect ratio of the bubble, $h_{max} / R$, in terms of the coefficients $c_1$ and $c_2$, parameters $Y$ and $\gamma$, and an external strain, $\epsilon$:
\begin{align}
\left( \frac{h_{max}}{R} \right)^2 &= - \frac{c_2 \epsilon}{5 c_1} + \sqrt{\left( \frac{c_2 \epsilon}{5 c_1} \right)^2 + \frac{\pi \gamma}{5 c_1 Y}}
\label{hr2}
\end{align}

In the absence of external strain, $\epsilon = 0$, this expression reduces to
\begin{align}
\frac{h_{max}}{R} &= \left( \frac{\pi \gamma}{5 c_1 Y} \right)^{1/4}
\label{hr3}
\end{align}
i.e., the value of $h_{max} / R$ is determined solely by the balance between vdW and elastic energies of a 2D crystal, independent of the properties of the substance captured within the bubble. This result is in excellent agreement with the constant aspect ratios observed experimentally - see Figs. \ref{fig:bubbles_scaling_graphene_MoS2_round} and \ref{fig:bubbles_scaling_graphene_trapez_smooth}.

The presence of finite $\epsilon$ (induced, for example, during fabrication) should modify the bubbles' shape, reducing the aspect ratio $h_{max}/R$ for tensile strains and increasing it for compressive strains - see Supplementary Note 2.

The above analysis also shows that the fluid material inside the bubble is under a constant hydrostatic pressure $P$, which is described by equations (6) and, following ref.$^{11}$, is referred to below as vdW pressure. Accordingly, our case of bubbles formed by the competition of vdW and elastic forces can be considered as a particular case of the membrane deformed by applying a constant external pressure.

Using the change of variables, $x = r / R$, we write the total energy as

\begin{align}
E_{tot} &= E_{el} + E_{bend} + P \times V \nonumber \\
E_{el} &= c_1 \left[ \tilde{h} ( x ) \right] Y \frac{h_{max}^4}{R^2} + c_2 \left[ \tilde{h} ( x ) \right] Y \epsilon h_{max}^2
\nonumber \\
E_{bend} &= c_3 \left[ \tilde{h} ( x ) \right] \kappa \frac{h_{max}^2}{R^2} \nonumber \\
E_{P} &= c_V \left[ \tilde{h} ( x ) \right] P h_{max} R^2
\end{align}

\begin{figure}
	\includegraphics[width=0.3\textwidth]{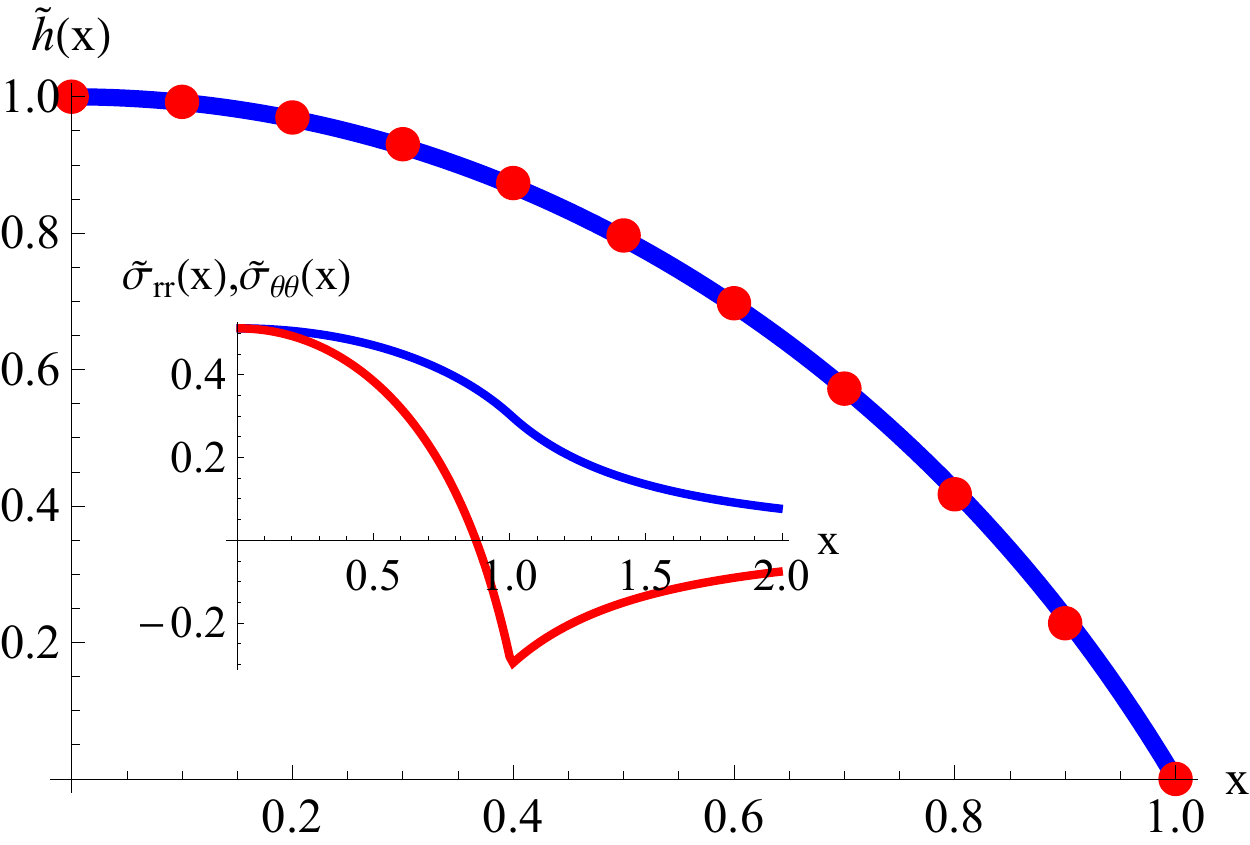}
	\includegraphics[width=0.3\textwidth]{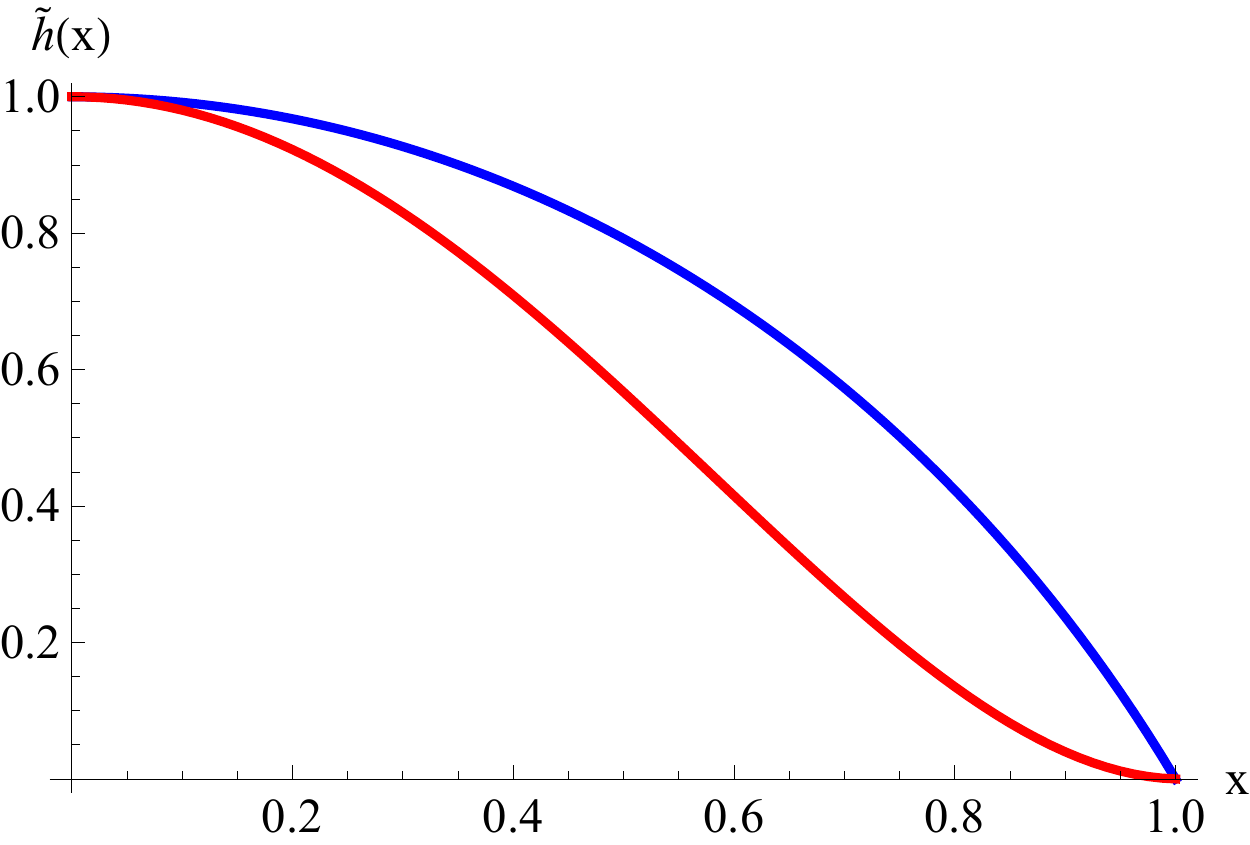}
	\caption{\label{fig:shape}
		\textbf{ Bubble profiles.} \textbf{Top panel}: Scaling function, $\tilde{h} ( x )$, obtained by minimizing numerically the elastic energy. It is well approximated by a quartic function, $\tilde{h} (x ) = 1 - x^2 + c ( x^2 - x^4 )$ (red dots), with $c = 0.25$. The inset shows the scaled stresses, $\tilde{\sigma}_{rr} ( x ))$ and $\tilde{\sigma}_{\theta \theta} ( x )$ (blue and red curves, respectively). \textbf{Bottom panel}: Comparison between the bubble profiles under a hydrostatic pressure for the cases dominated by in--plane strains (blue) and bending (red).}
\end{figure}

We consider first the bubble's profile, $\tilde{h} ( x )$, determined solely by the competition between the pressure and the in-plane stresses, $E_{el}$ and $E_P$, and we set $\epsilon = 0$. Minimization of $E_{tot}$ with respect to $h_{max}$ gives
\begin{align}
h_{max} &= \left[ \frac{c_V ( \tilde{h} )}{c_1 ( \tilde{h} )} \right]^{1/3} \left( \frac{P R^4}{4 Y} \right)^{1/3} \nonumber \\
E_{tot} \left[ \tilde{h} \right]&= - \frac{3}{4} \left[ \frac{c_V^4 ( \tilde{h} )}{c_1 ( \tilde{h} )} \right]^{1/3} \left( \frac{P^4 R^4}{4 Y} \right)^{1/3}
\end{align}
We can now calculate $\tilde{h}$ by minimizing $E_{tot}$. This yields that $\tilde{h} ( x )$ is universal, i.e. independent of $Y, P$ and $R$. The function $\tilde{h} ( x )$ is shown in Fig. \ref{fig:shape}. The in-plane stresses associated with the bubble formation can also be expressed in a scaled form, $\tilde{\sigma}_{rr} ( x ) = [ R^4 / ( h_{max}^4 Y ) ] \times \sigma_{rr} ( r / R )$ and $\tilde{\sigma}_{\theta \theta} ( x ) = [ R^4 / ( h_{max}^4 Y ) ] \times \sigma_{\theta} ( r / R )$. These functions are plotted in the inset of Fig. \ref{fig:shape}. It is interesting to note that the hoop stress, $\sigma_{\theta \theta}$, becomes negative (compressive) near the base of the bubble. In the absence of vdW pressure, a compressive stress can lead to an instability with respect to the formation of wrinkles$^{19}$. The existence of in-plane stresses outside the bubble (see inset if Fig. \ref{fig:shape}) implies that bubbles interact with each other - see Supplementary Note 2 where this interaction is analyzed. It is attractive and decays as $Y / d^2$, where $d$ is the distance between bubbles.

A similar analysis can be carried out when the shape of the bubble is determined by the bending rigidity, and $E_{tot} = E_{bend} + E_{P}$. In this case we find
\begin{align}
h_{max} &= \frac{c_V ( \tilde{h} )}{2 c_3 ( \tilde{h} )} \frac{P R^4}{\kappa} \nonumber \\
E_{tot} \left[ \tilde{h} \right]&= - \frac{c_V ( \tilde{h} )}{4 c_3 ( \tilde{h} )} \frac{P^2 R^6}{\kappa}
\end{align}
The generic profiles in the two cases (elastic energy is dominated by either in-plane stresses or bending) are given in Fig. \ref{fig:shape}.

In the following, we neglect the bending rigidity term, $E_{bend}$, as appropriate for 2D membranes with $\sqrt{\kappa / Y} \ll h_{max} , R$, and corresponds to the case studied in our experiments. In principle, the coefficients $c_1$ and $c_2$, and the function $\tilde{h}$, can depend on strain, $\epsilon$. However, we have found numerically that this dependence is negligible for $\epsilon \lesssim 0.1$, i.e., can be neglected in realistic situations because even smaller strains (a few \%) are likely to cause slippage along the substrate due to limited adhesion. The numerical parameters that relate $h_{max} , L , Y$ and $P$ are found as
\begin{align}
c_1 &\approx 0.7 \nonumber \\
c_2 &\approx 0.6 \nonumber \\
c_V &\approx 1.7
\end{align}

\begin{figure}
	\includegraphics[width=0.5\textwidth]{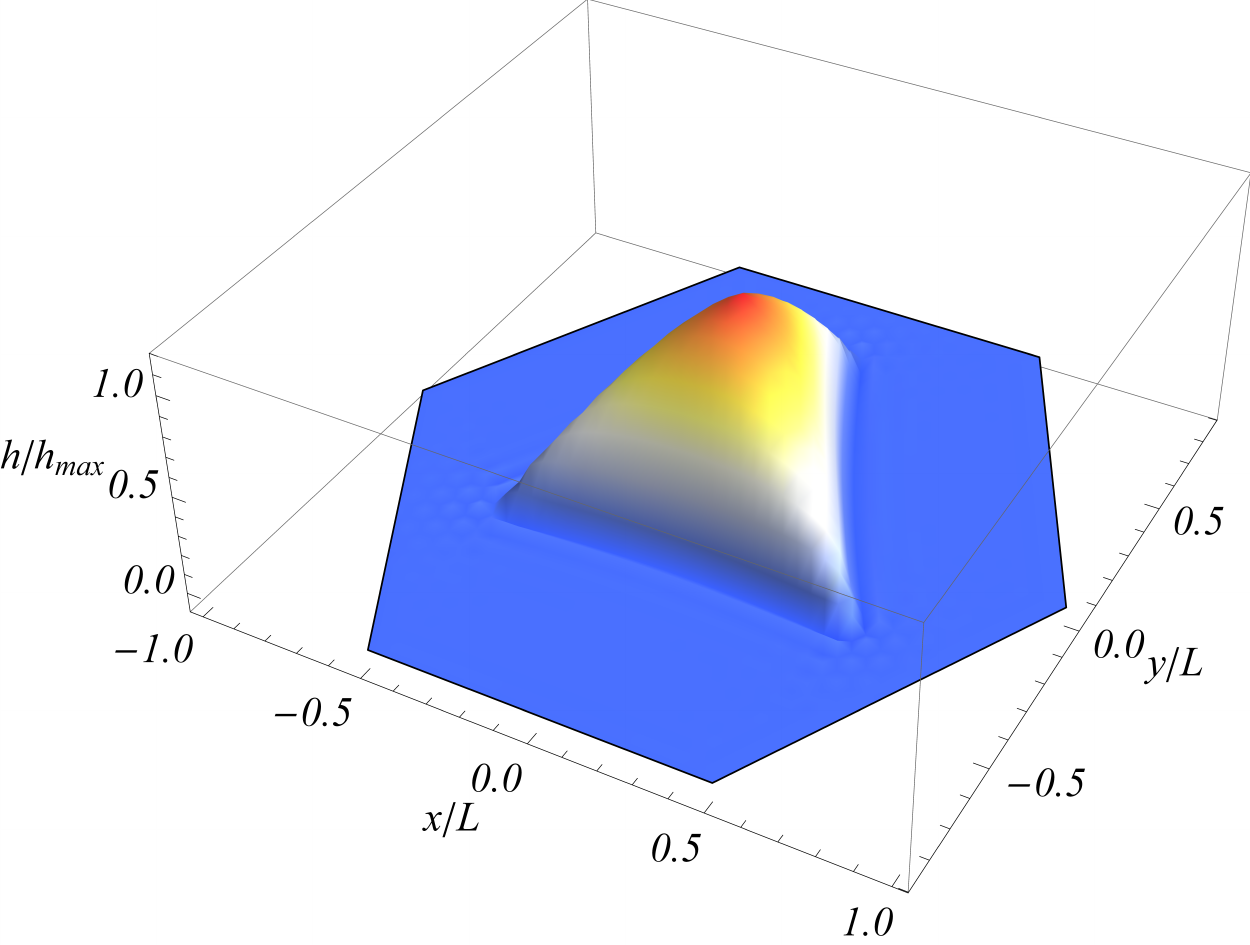}
	\caption{\label{fig:triangle}
		\textbf{Triangular bubble.} Universal shape of triangular bubbles with the equilateral base (theory).}
\end{figure}

The above scaling analysis can also be applied to bubbles of other shapes, such as the pyramidal bubbles found experimentally (Fig. \ref{fig:bubbles_scaling_graphene_trapez_smooth}). For simplicity, we model smooth triangular bubbles as having an equilateral triangle as their base. The bubbles are then characterised by two length scales: height, $h_{max}$, and the side length, $L$. The scaled universal profile for a triangular bubble is shown in Fig. \ref{fig:triangle}. The numerical parameters in this case are
\begin{align}
c_1 &\approx 0.6 \nonumber \\
c_2 &\approx 0.3 \nonumber \\
c_V &\approx 0.2.
\end{align}

The corresponding average strain, $\bar{u}_{rr}$, for graphene/hBN/MoS$_2$ monolayers enclosing a bubble is of order
\begin{align}
\bar{u}_{rr} &\approx \left( \frac{h_{max}}{R} \right)^2 \approx 1 - 2\%
\label{strain}
\end{align}

To gain further insight, we estimate the parameters in equation (\ref{hr3}) corresponding to the experimentally observed aspect ratio $h_{max} / R \approx 0.11$ for circular and $h_{max} / L \approx0.07$ for triangular graphene bubbles. Using the known stiffness of graphene, $Y_G \approx 22$ eV \AA$^{-2}$, this yields an effective adhesion energy $\gamma\sim0.005$ eV \AA$^{-2}$, significantly lower than the measured value for adhesion between graphene on SiO$_x$, $\sim 0.03$ eV \AA$^{-2}$  (ref. 20) and, also, lower than the vdW adhesion found theoretically$^{21,22}$, $\sim 0.01-0.02$ eV \AA$^{-2}$. This indicates that the adhesion between graphene (or hBN, MoS$_2$) and the trapped hydrocarbon contamination, $\gamma_{Gb}$, is comparable to that between graphene and the substrate, $\gamma_{GS}$, as expected for these lipophilic 2D crystals. Note that $\gamma_{Gb}$ should be smaller than $\gamma_{GS}$. Otherwise, no bubbles would be formed as the contaminating materials would tend to spread along the substrate. As we show below, a similarly low $\gamma$ follows from our AFM measurements of vdW pressure inside bubbles (see Section \ref{sec:pressure}).

hBN has approximately the same stiffness as graphene$^{23}$ which results in similar aspect ratios because they depend only weakly of $Y$ (as $Y^{1/4}$). On the other hand, MoS$_2$ is significantly less stiff, with twice lower Young's modulus $Y_{MoS_2} \approx 11.2$ eV \AA$^{-2}$  (refs. 15,16,24). This translates into a larger aspect ratio compared to graphene and hBN, in agreement with the experiment.
Furthermore, notably different aspect ratios for MoS${_2}$ bubbles on hBN and MoS$_2$ substrates ($\approx0.14$ vs $\approx0.17$, see Fig. \ref{fig:bubbles_scaling_graphene_MoS2_round}b) can be attributed to different $\gamma$ for the two substrates; see equation (\ref{hr3}).

\begin{figure}
	\includegraphics[width=0.5\textwidth]{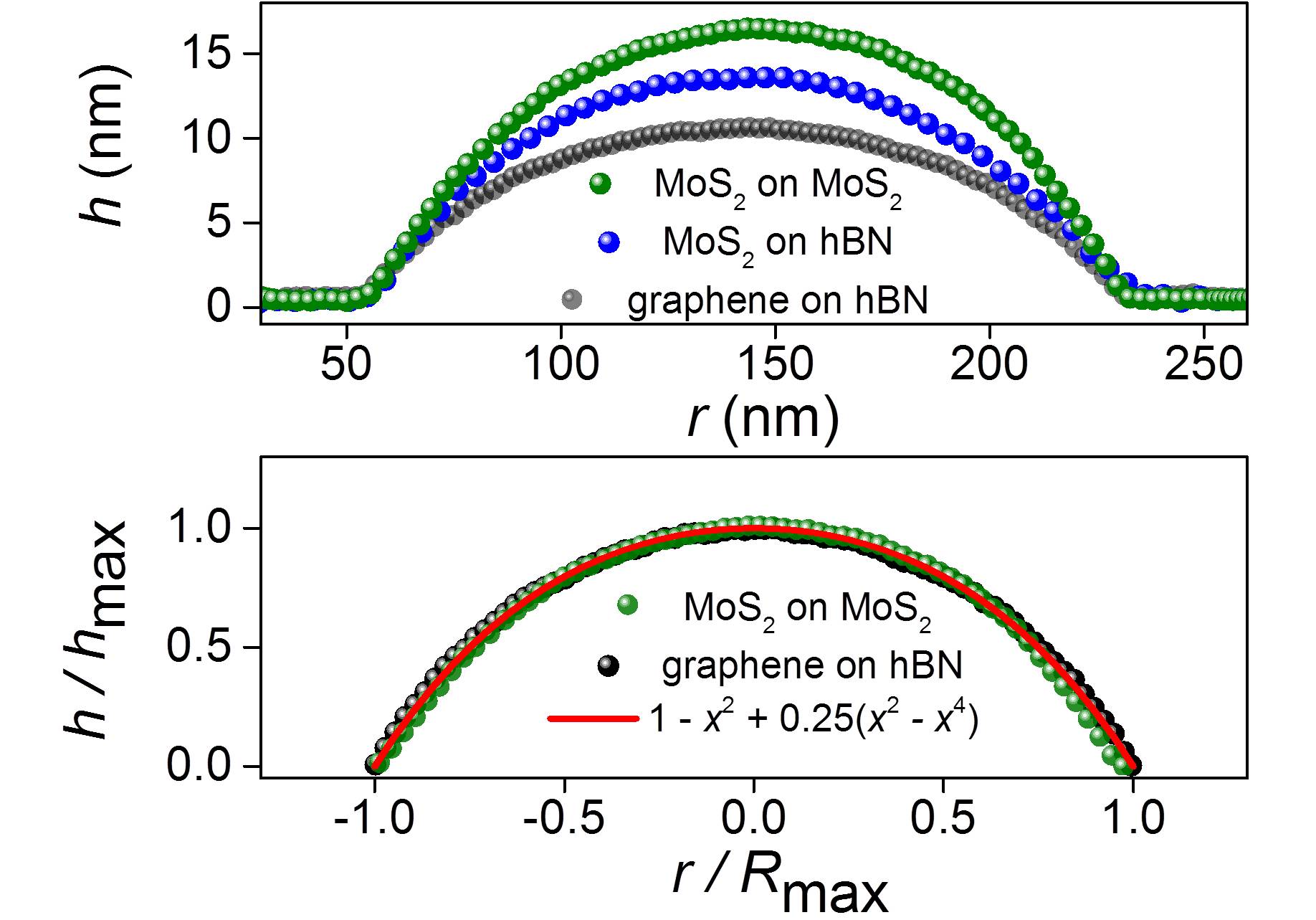}
	\caption{\label{fig:G_MoS2_BN_profile}
		\textbf{Universal profile of round bubbles}.
		Scaled and normalized cross-sectional profiles measured experimentally for typical round bubbles formed by graphene and MoS$_2$ monolayers. \textbf{Top panel:} Comparison of the profiles of graphene and MoS$_2$ bubbles scaled to the same radius, $R$, in order to emphasize their different heights. The actual sizes are $R = 168$ nm for graphene and $R = 97$ and $816$ nm for MoS$_2$ monolayers on hBN and MoS$_2$, respectively. \textbf{Bottom panel:} Same data but scaled in both dimensions. The symbols show the experimental profiles; the red curve is the theory profile as in Fig. \ref{fig:shape}a, with no fitting parameters.}
\end{figure}

Fig. \ref{fig:G_MoS2_BN_profile} compares the calculated universal profile, $\tilde{h} ( x )$, with those observed experimentally for the round bubbles formed by graphene and MoS$_2$ monolayers. In both cases the profiles are remarkably well described by the quartic function shown in Fig. \ref{fig:shape}a.
This proves that not only the aspect ratio, $h_{max}/R$, but also the shape of the bubbles is universal and determined solely by the elastic properties of 2D crystals and their adhesion, independent of the properties of the trapped material.

\section{Deviations from scaling}
\label{sec:deviations}
In the experiments, the predicted scaling behavior breaks down for large pyramidal bubbles with sharp ridges, pointed summits and relatively flat facets, such as those shown in Fig. \ref{fig:bubbles_scaling_graphene_trapez}. They exhibit a significant spread of $h_{max}/L$ values, from $\sim$ 0.09 to 0.2, which are also larger than the values found for smooth bubbles (Fig. \ref{fig:bubbles_scaling_graphene_trapez_smooth}).

Sharp ridges between flat facets minimize the in-plane elastic energy at the cost of bending along the length of the ridge$^{25}$. Therefore
we assume that most of the elastic energy of such bubbles resides in the ridges. Following the analysis in ref.$^{25}$ we consider a ridge of length $L$ separating two flat facets that make an angle $\theta$. For a bubble of height $h_{max}$ with a pyramidal shape, where the basis is a polygon with the side length $L$, we have $\theta \approx h_{max} / L$. At the center of the ridge, its curvature can be described by the radius $R_{ridge}$ and, in order to allow for that curvature to exist, the ridge has to sag by an amount $\xi \sim R_{ridge} \theta^2$. The strained area around the ridge has a width $w \sim R_{ridge} \theta$ and a length $L$. Then, the resulting in-plane elastic energy is of the order of
\begin{align}
E_{str} &\approx Y w L \left( \frac{\xi}{L} \right)^2 \approx Y \frac{R_{ridge}^5}{L^3} \theta^9
\end{align}
The associated bending energy scales as
\begin{align}
E_{bending} &\approx \kappa \frac{w L}{R_{ridge}^4} \approx \kappa \frac{L}{R_{ridge}} \theta
\end{align}
The optimal value of $R_{ridge}$ makes these two energies comparable, so that
\begin{align}
R_{ridge} &\approx \left( \frac{L}{\theta^2} \right)^{2/3} \left( \frac{\kappa}{Y} \right)^{1/6}
\end{align}
and the total elastic energy is of the order of
\begin{align}
E_{str} + E_{bending} &\approx \kappa L^{1/3} \theta^{5/3} \approx \kappa \left( \frac{Y}{\kappa} \right)^{1/6} \frac{h_{max}^{5/3}}{L^{4/3}} \left( \frac{Y}{\kappa} \right)^{1/6}.
\end{align}
The relation between $h_{max}$ and $L$ is given by the minimization of the elastic and vdW energies, where, as in the previous section, $E_{vdW} \propto \gamma L^2$. We finally find
\begin{align}
\frac{h_{max}^{5/3}}{L^{10/3}} &\propto \frac{\gamma}{\kappa^{5/6} Y^{1/6}} \nonumber \\
\frac{h_{max}}{L} &\propto L \frac{\gamma^{3/5}}{\kappa^{1/2} Y^{1/10}} =  \frac{L}{L_0}
\end{align}
with
\begin{align}
L_0 &= \frac{\kappa^{1/2} Y^{1/10}}{\gamma^{3/5}},
\end{align}
i.e., the aspect ratio of bubbles with sharp ridges is not constant but depends on the size and geometry of the bubbles, which explains the absence of a universal scaling in this case as observed experimentally.

\begin{figure}
	\includegraphics[width=0.48\textwidth]{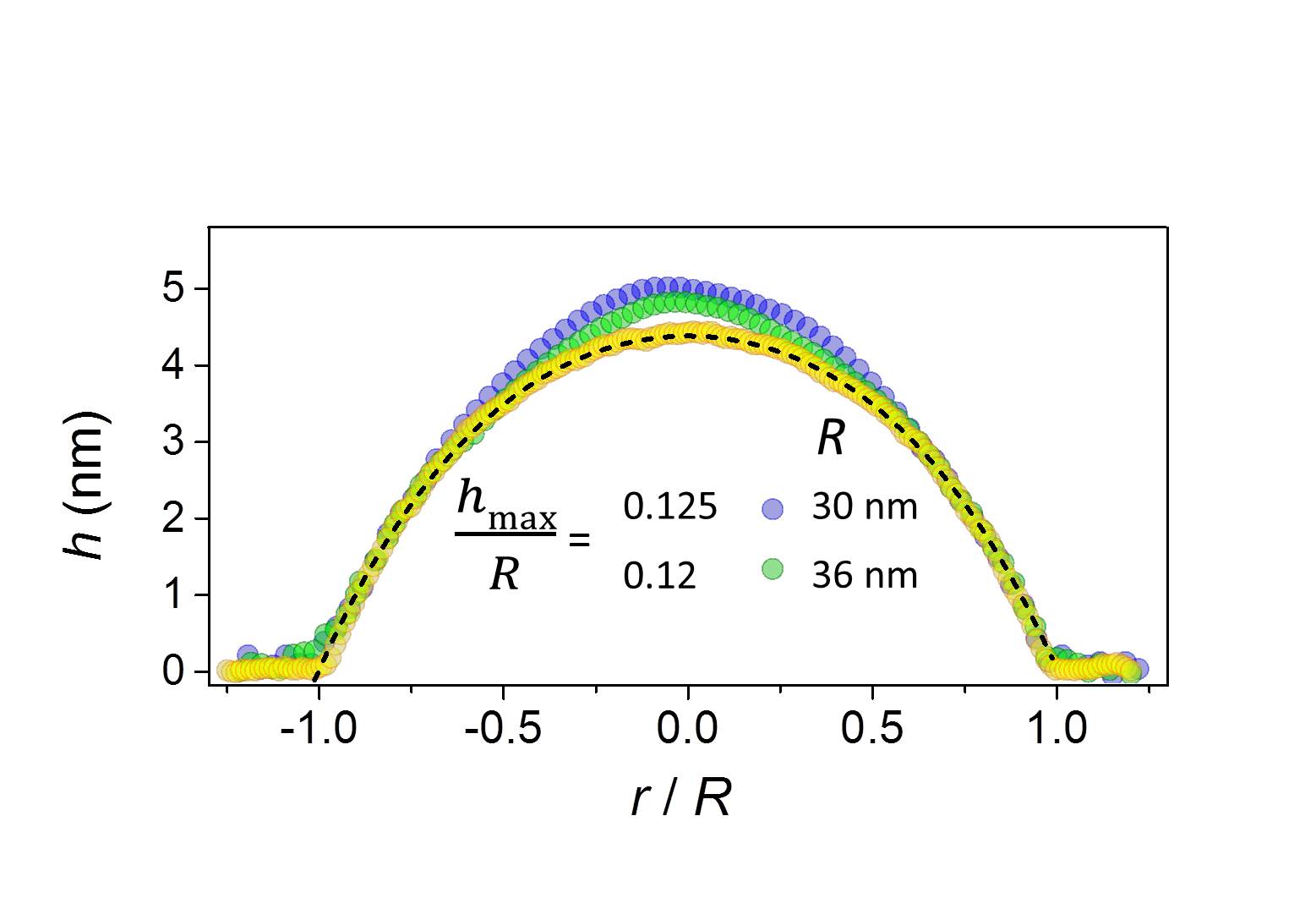}
	\caption{\label{fig:small_bubbles_profiles}
		\textbf{Deviations from the universal profile in small bubbles.}
		Scaled experimental profiles of small graphene bubbles showing deviations (blue and green symbols). For comparison, the universal profile of a larger graphene bubble is shown by yellow symbols, together with the fit to the theoretical scaling function (dashed curve).}
\end{figure}

Deviations from the universal profile were also found for very small graphene  and hBN bubbles, $R \lesssim 50$nm, despite the fact that they seem to be smooth and almost perfectly round (see Fig. \ref{fig:bubbles_scaling_graphene_MoS2_round}a). We attribute the breakdown of scaling in this case to residual strain in the 2D layer, i.e., unlike large round bubbles, small ones were not fully relaxed during their annealing. This is consistent with the bubbles' profiles observed in the two cases - see Fig. \ref{fig:small_bubbles_profiles}. While all larger bubbles exhibited the universal profile described by $\tilde{h} (x ) = 1 - x^2 + c ( x^2 - x^4 )$ (yellow dots in Fig. \ref{fig:small_bubbles_profiles}), the small ones showed notable deviations near the top (blue and green dots in Fig. \ref{fig:small_bubbles_profiles}), indicating some residual compressive strain. The latter favors higher values of $h_{max} / R $; see Supplementary Note 2. Using this Note's equations, we estimate that the observed deviations in small bubbles would require a compressive strain $| \epsilon |$ of the order of $\lesssim 10^{-3}$, in good agreement with remnant strains usually observed by Raman spectroscopy$^{26,27}$.

\section{Pressure inside the bubbles.}
\label{sec:pressure}
Equation (\ref{min}) allows us to calculate the pressure inside a bubble:
\begin{align}
P &= \frac{Y}{c_V h_{max}} \left[ 4 c_1 \left( \frac{h_{max}}{R} \right)^4 + 2 c_2 \epsilon \left( \frac{h_{max}}{R} \right)^2 \right]
\label{pr}
\end{align}
Using eqs. (\ref{hr2}) and (\ref{hr3}), we find, for $\epsilon = 0$,
\begin{align}
P &= \frac{4 \pi \gamma}{5 c_V h_{max}},
\label{press}
\end{align}
i.e., vdW pressure is determined by the adhesion between a 2D crystal and the substrate and their separation. This expression is similar to the estimate given in ref. 11.

To find the dependence of $P$ on the bubble volume, $V$, we write $h_{max}$ as a function of $V$ and obtain
\begin{align}
P &= \frac{4 \pi \gamma}{5} \left( \frac{5 c_1 Y}{\pi \gamma} \right)^{1/6} \left( \frac{c_V}{V} \right)^{1/3}
\label{pressv}
\end{align}
The pressure is independent of the compressibility of the material within the bubble, i.e., $V$ adjusts itself in such a way that the pressure exerted on the material inside it (or, vice versa, acting on a 2D membrane) has the value required by the equilibrium between vdW and elastic forces.

If a gas is trapped inside, it's compressibility depends on temperature, and, for a monoatomic gas,  $P = N k_B T / V$. This relation, together with equation (\ref{hr3}), implies
\begin{align}
h_{max} &= \left( \frac{N k_B T}{c_V P} \right)^{1/3} \left( \frac{\pi \gamma}{5 c_1 Y} \right)^{1/6}
\end{align}
and, combined with equation (\ref{press}), gives
\begin{align}
P &= \left( \frac{4 \pi \gamma}{5} \right)^{3/2} \left( \frac{c_V}{N k_B T} \right)^{1/2} \left( \frac{5 c_1 Y}{\pi \gamma} \right)^{1/4}
\label{presst}
\end{align}
For a 1 $\mu$m$^3$ volume of a gas captured under ambient conditions (1 atm at room $T$) between a substrate and a 2D membrane, the gas would be compressed to about 1\% of its initial volume and experience $P$ of $\approx 4$  MPa. This implies a density of $2\times10^{20}$ cm$^{-3}$, which is likely to turn many gases (including water vapor) into liquids, and the preceding analysis [equations (\ref{pr}-\ref{pressv})] then becomes more appropriate.

\begin{figure}
	\includegraphics[width=0.42\textwidth]{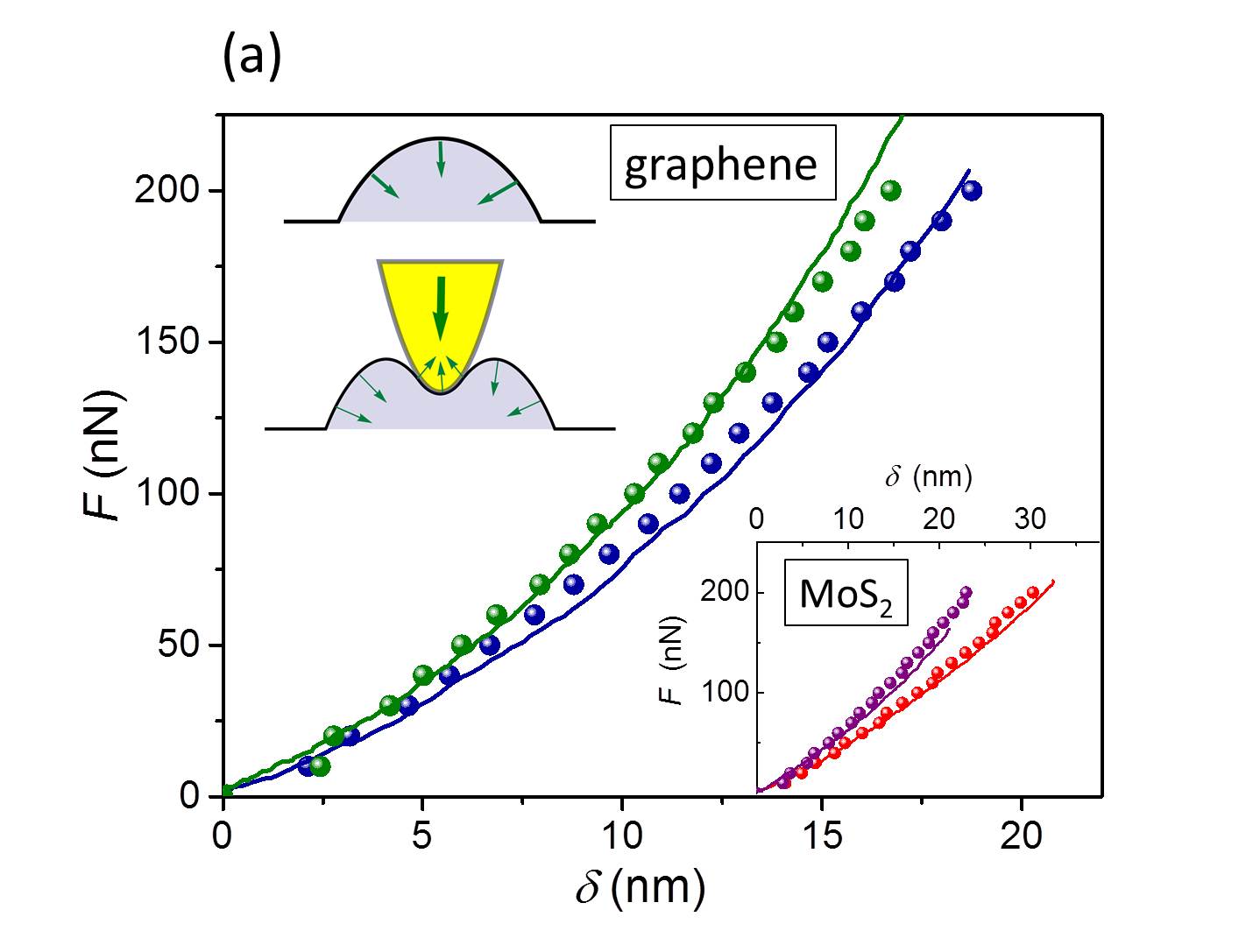}
	\includegraphics[width=0.43\textwidth]{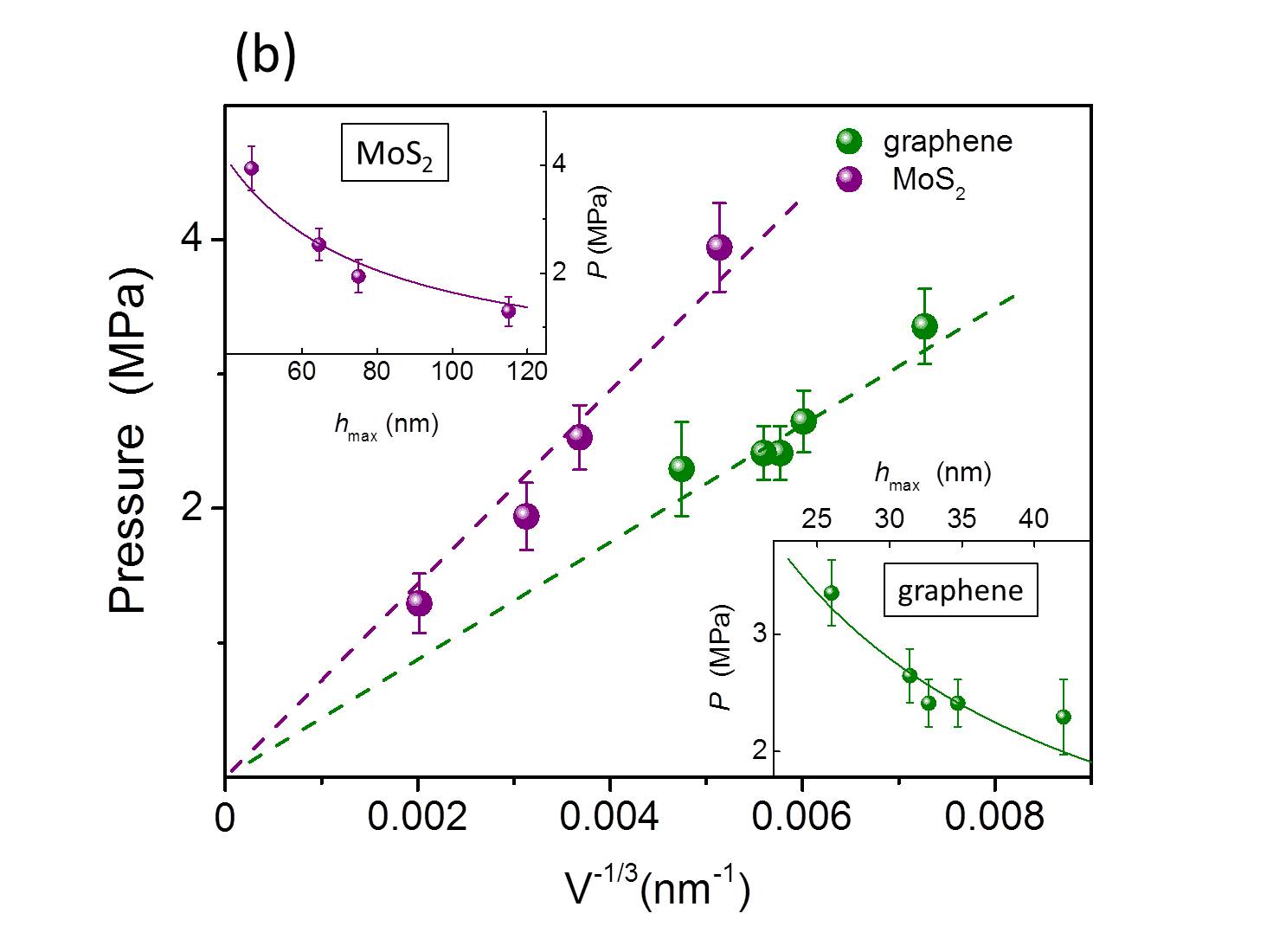}
	\includegraphics[width=0.45\textwidth]{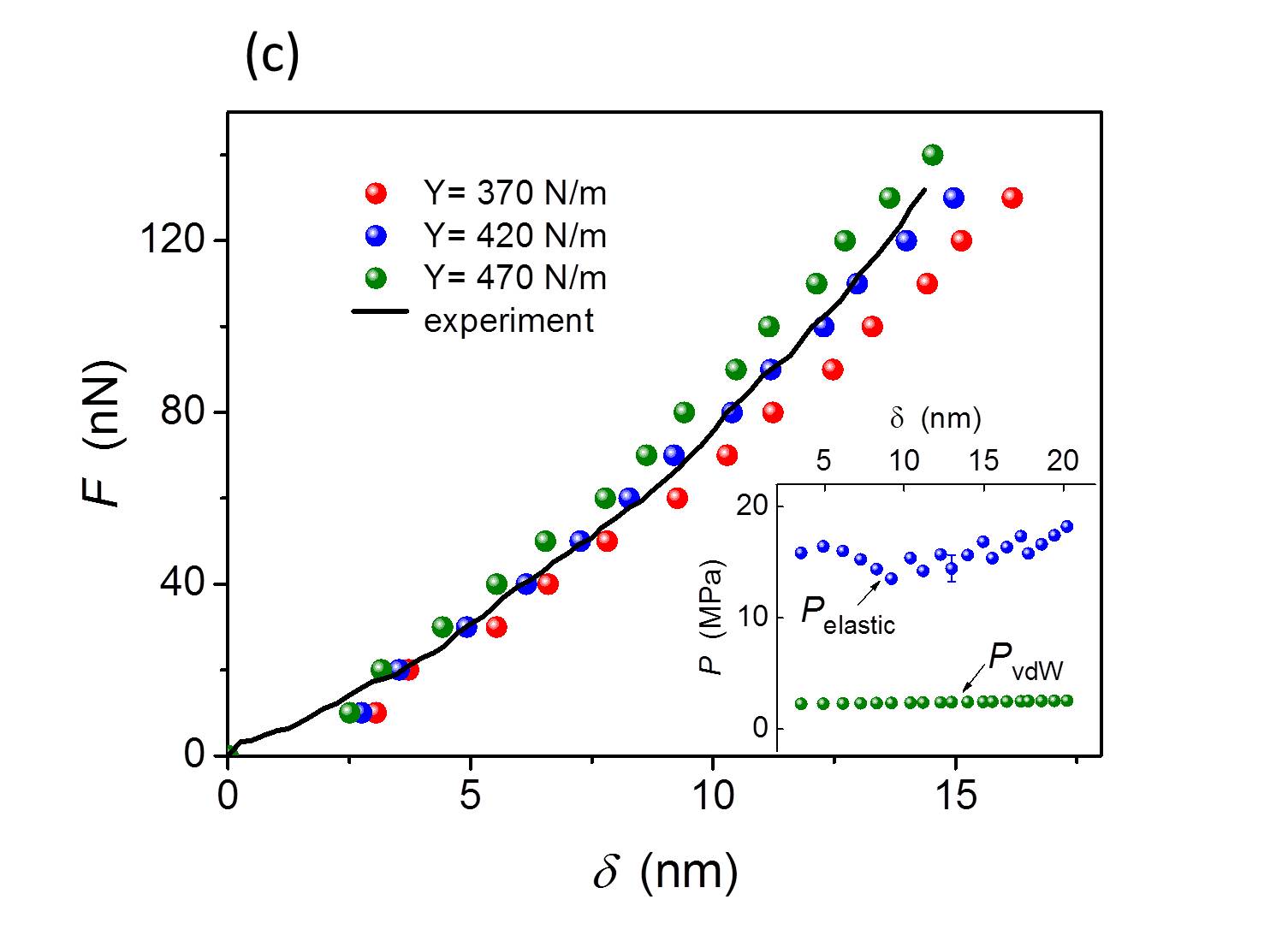}
	\caption{\label{fig:pressure}
		\textbf{Measuring vdW pressure}. \textbf{(a)} Main panel: experimental FDCs (solid curves) and numerical fits (symbols) for two graphene bubbles of different sizes ($h_{max}= 26$ and 33 nm; $R=240$ and 320 nm, respectively). Top left inset: Sketch of the forces exerted on an AFM tip and a material inside the bubble before (top) and during (bottom) indentation. Bottom-right inset: experimental FDCs (solid) and numerical fits (symbols) for two monolayer MoS$_2$ bubbles ($h_{max}= 64.5$ and 115 nm; $R=437$ and 800 nm, respectively). \textbf{(b)} Main panel: vdW pressure in graphene and MoS$_2$ bubbles as a function of their volume. Dashed lines are linear theory fits to the data. Top-left and bottom-right insets show the same pressure data but as a function of the maximum height. Solid lines are fits to $\propto 1/h_{max}$. \textbf{(c)} Main panel: An experimental FDC (black curve) and numerical fits for three different values of Young's modulus. Inset: Comparison of the typical values of pressure exerted on the AFM tip due to the elastic deformation of graphene, $P_{elastic}$,  and of the vdW pressure inside the bubble, $P_{vdW}$. Shown are data for one of the graphene bubbles ($h_{max}$= 33 nm and $R$ =320 nm).}
	
\end{figure}

To measure the pressure inside the observed bubbles, we used nanoindentation with an AFM tip, an approach similar to that used e.g. in ref. 28 to measure the osmotic pressure inside viral particles. We indented bubbles of different sizes with an AFM tip and recorded their force-distance curves (FDCs) (see Methods for details). To ensure a smooth spherical shape of the used AFM tips, they were annealed at a high temperature (Supplementary Fig. S1). Typical FDC's for several graphene and MoS$_2$ bubbles are shown in Fig. \ref{fig:pressure}a. One can see that, as the bubble size decreases, the force, $F$, required to achieve a certain indentation depth, $\delta$, increases. This is qualitatively consistent with the expectation that the vdW pressure should increase as $1/h_{max}$ (see equations (\ref{press},\ref{pressv})). However,
the force measured by a nanoindentation probe results not only from the resistance due to a finite pressure inside a bubble but also from the accompanying elastic deformation of the 2D membrane, as we show next.

To separate these two contributions, we have analyzed the total energy of a pressurized bubble subject to indentation from an approximately spherical AFM tip. The forces acting on the enclosed material and on the AFM tip are sketched in the inset of Fig. \ref{fig:pressure}a. In the absence of indentation, graphene induces a downward pressure described by equations (\ref{pr}--\ref{pressv}). As the AFM tip starts to create a dent at the top of the bubble, the resulting additional deformation of graphene creates a pressure in the opposite direction that partially compensates the pressure exerted by the tip. The sum of the two pressures is equal to the pressure from the AFM tip:

\begin{align}
P_{tip} = F(\delta)/S(\delta)
\label{afm}
\end{align}
where $\delta$ is the indentation depth, $F(\delta)$ the applied force and $S(\delta)$ the part of the tip surface area in direct contact with the bubble. Therefore, the vdW pressure, $P$, can be found as a difference between $P_{tip}$ measured experimentally and the elastic energy contribution that we evaluate below.

The total energy of the bubble, for an indentation $\delta$, can be written as a sum of the work done by the force $F$ and the elastic, vdW, and internal energies:
\begin{align}
E_{tot} &= - F \delta + E_{el} ( R , \delta ) + E_{vdW} ( R ) + E_V [ V ( R , \delta ) ]
\end{align}
where $E_{vdW}$ is given by equation (\ref{EvdW}).
We assume that the energy of the material inside the bubble can be written in terms of its volume, $V ( R , \delta )$, only. The elastic energy can be written as
\begin{align}
E_{el} ( R , \delta ) &= E_{el}^0 ( R ) + \delta E_{el} ( R , \delta )
\end{align}
where $E_{el}^0 ( R )$ is the elastic energy before indentation.

To estimate the effect of indentation on $\delta E_{el} ( R , \delta )$ for small indentations, $\delta \ll h_{max}$, we describe graphene at the top of the bubble as an almost flat membrane under a uniform tensile stress, $\sigma$. Given the small aspect ratios of all our bubbles, $h_{max}/R < 0.15$, this assumption is justified  both for our graphene and MoS$_2$ membranes. Then the indentation deforms the bubble over a region of radius $R^* \lesssim R$. On dimensional  grounds, the elastic energy due to the indentation can be written as
\begin{align}
\delta E_{el} ( \delta , R ) &=  \sigma \int_0^{R^*} 2 \pi r d r \left( \frac{\partial h}{\partial r} \right)^2 = c ( \nu )  \sigma \delta^2
\end{align}
where $c ( \nu )$ is a numerical constant that depends on the Poisson ratio of the membrane, $\nu$. As $\delta E_{el}$ does not depend on $R^*$ or $R$, the minimization of $E_{tot} ( R , \delta )$ with respect $R$ leads to
\begin{align}
0 &= \frac{\partial E_{el}^0 ( R )}{\partial R} + \frac{\partial E_V ( V )}{\partial V} \frac{\partial V}{\partial R} + 2 \pi \gamma R
\end{align}

This equation defines the dependence $R ( \delta )$ of the bubble radius on the indentation depth. As a hydrocarbon material inside bubbles is essentially incompressible, we assume that $V [ R ( \delta ) , \delta ]$ does not change. Then, the minimization of the total energy with respect to $\delta$ yields
\begin{align}
F &= 2 c ( \nu ) \sigma \delta,
\end{align}
i.e., the relation between the force and the indentation depth is expected to be linear. The nonlinearity of FDCs observed experimentally (see Fig. \ref{fig:pressure}) arises due to a nonlinear dependence of the contact area between the AFM tip and the pressurized bubble on $\delta$.

The value of $\sigma$ is determined by $Y$, and the strain $\epsilon$ that scales as
\begin{align}
\epsilon &\propto \frac{h_{max}^2}{R^2}
\end{align}
which leads to
\begin{align}
\frac{F}{\delta} &= d ( \nu ) Y \frac{h_{max}^2}{R^2}
\label{indent}
\end{align}
where $d ( \nu )$ is another dimensionless constant which depends on the Poisson ratio. If the force is applied over a finite area, defined by a contact radius $R_{contact}$,  the value of $d$ in equaion (\ref{indent}) also depends on the ratio $R_{contact} / R$.

The vertical pressure due to the elastic deformations can be written as
\begin{align}
P_{el} ( r ) &= \frac{1}{r} \frac{\partial}{\partial r} \left[ r  \sigma_{rr} ( r ) \partial_r h ( r )\right]
\label{el}
\end{align}
where $\sigma_{rr} ( r )$ is the radial stress. For a flat (e.g. cylindrical) tip of a radius comparable with the radius of a bubble, one would have $\partial_r h ( r ) = 0$ and $P_{el} = 0$, so that the pressure in the contact area would be simply $P = P_{tip} = F_{tip} / A_{tip}$, i.e. the AFM tip would directly measure the pressure inside the bubbles, as in Imbert-Fick tonometry law$^{29}$. In our case, the tip is spherical, the contact radius, $R_{contact}$, is significantly smaller than $R$ and depends on $\delta$, which results in an additional contribution from elastic forces.

We have used equations (26-29) to numerically fit the experimental FDCs for several graphene and MoS$_2$ bubbles, taking into account changes in $R_{contact}$ with $\delta$, as well as the increase in bubble's radius during indentation. The extracted values of vdW pressure and their dependence on $h_{max}$ and $V$ are shown in Fig. \ref{fig:pressure}b. For $R$ in the range 250--800 nm and heights $h_{max} = 26 - 115$ nm, graphene and MoS$_2$ monolayers exert $P$ of the order of several MPa, or tens of bar. The $P(h_{max})$ and $P(V)$ dependences obtained from experimental FDCs are in good agreement with our theory (equations 23--24) -- see Fig. \ref{fig:pressure}b. The agreement implies that for graphene bubbles of a smaller height ($h_{max} \sim 1$ nm) $P$ can easily reach $\sim 100$ MPa. This is somewhat lower than the 1 GPa estimate given in ref. 11 for a hydrophobic material captured inside graphene bubbles. To this end, let us recall that $\gamma$ in equations (17-18) is the difference between graphene's adhesion to the substrate, $\approx 30$  meV \AA$^{-2}$ (ref. $^{20}$) and its adhesion to a material inside bubbles (equation (\ref{EvdW})). As graphene is lipophillic, its adhesion to hydrocarbons can be expected to be significant, thus reducing the effective $\gamma$. From our data we extract $\gamma_{graphene} = 3.8 \pm 0.3$  meV \AA$^{-2}$ and $\gamma_{MoS_2} = 6.8 \pm 0.6$  meV \AA$^{-2}$ for bubbles enclosing hydrocarbons. The twice higher value of $\gamma_{MoS_2}$ compared to graphene corresponds to higher vdW pressures for the same bubble size (see Fig. \ref{fig:pressure}b).

In addition to vdW pressure, our indentation experiments allowed us to estimate the elastic stiffness (Young's moduli) of the studied 2D membranes. Fig. \ref{fig:pressure}c illustrates that our numerical fits to experimental FDC's are very sensitive to $Y$: changing its value in numerical fitting by 5--10\% allowed us to narrow down the value of Young's modulus to $Y_{graphene} = 420 \pm 20$ N/m and $Y_{MoS_2} = 210 \pm 20$ N/m. Both values are somewhat higher than the reported average values of $Y$ obtained using nanoindentation of suspended membranes, $350 \pm 50$ N/m for graphene$^{30}$ and $180 \pm 80$ N/m for MoS$_2$ (refs.$^{15,16,24}$) . This can be due to the fact that our 2D membranes are strained by $\sim 1\%$ due to high pressure inside. This can increase their stiffness as suggested recently$^{31,32}$.

Let us also note that, due to the high stiffness of our 2D membranes, the elastic contribution to the measured force acting on the AFM tip is comparable with that due to vdW pressure - see inset in Fig. \ref{fig:pressure}c. Both pressures are approximately constant, i.e., independent of the indentation depth, as expected. (The apparent variations in $P_{elastic}$ are due to the discreet nature of our numerical fitting: its value is sensitive to details of the contact between the AFM tip and bubble, which cannot be accurately reproduced at each value of $\delta$.) This implies that analysis of nanoindentation experiments in the presence of hydrostatic pressure must take into account both contributions, as done in our work.

\section{Conclusions}
We have shown that bubbles formed by monolayers of graphene, hBN and MoS$_2$ deposited onto atomically flat substrates exhibit the universal behavior determined purely by elastic properties of the 2D crystals and independent of properties of a trapped material.

Bubbles with smooth shapes exhibit the same aspect ratio, $h_{max} / L_{eff}\sim 0.1$, independent of their size, where $h_{max}$ is the height of the bubble, and $L_{eff}$ is the characteristic length scale that describes their base. For round bubbles $L_{eff} = R$ and for triangular ones $L_{eff} \sim L$, and the distribution of values of $h_{max} / L_{eff}$ is quite narrow. The average strain in the 2D crystal enveloping such bubbles is $\sim ( h_{max} / L_{eff} )^2 \approx 10^{-2}$.

Our scaling analysis shows that the value of $h_{max} / L_{eff}$ is determined by competition between vdW adhesion and elastic energies. The vdW contribution favors the formation of bubbles with a small base, and the elastic energy tends to minimize their height. While $h_{max} / L_{eff}$ does not depend on the bubble size, it depends on whether or not any residual strains in the 2D crystals are present (for example, unrelaxed strains introduced during fabrication). The remnant strain contributes mostly to the shape of the smallest bubbles, with sizes below 50 nm.

Using AFM indentation, we were able to measure the vdW pressure exerted by graphene and MoS$_2$ membranes on the trapped material and to extract values of vdW adhesion and Young's moduli. The pressure can be approximated by $P \sim \gamma / L_{eff} \lesssim \gamma / h_{max}$. For the relatively large bubbles in our experiments ($R$ =250-800 nm) the measured vdW pressures were in the range 1.5-4 MPa, an order of magnitude lower than could be expected from the known adhesion energy between graphene and SiO$_x$. This is attributed to non-negligible adhesion between graphene and enclosed hydrocarbons. In situations where the adhesion between graphene and trapped substances is weak, as in the case of trapped water, the vdW pressure is expected to be much higher$^{11}$. It would be particularly interesting to measure the vdW pressure in true-nanoscale bubbles with $h\sim 1$ nm, which were not accessible in our experiments but can exhibit pressures of the order of 1 GPa.

The combination of topographic and indentation experiments on graphene and MoS$_2$ bubbles provides an excellent method to determine the materials' elastic properties.

\section*{Methods}

The AFM images and force-distance indentation curves$^{33}$ (FDCs) were obtained using Bruker Dimension FastScan AFM. The aspect ratio of the bubbles was measured in non-contact mode using soft cantilevers (nominal spring constant $k =0.7$ N/m; nominal tip radius $r = 2$ nm) in order to minimise tip-sample interaction and avoid modifying the shape of the bubbles.

Indentation experiments on both graphene and MoS$_2$ bubbles were performed using silicon probes with $k = 200$ N/m and $r = 8$ nm.
In order to increase the contact area between the tip and pressurized bubbles, the cantilevers were further annealed in air at 1000$^{\circ}$C for 2h (refs. 34,35). This treatment increased $r$ from the nominal 8 to $\sim100$ nm and imparted a smooth spherical shape, as shown by the SEM image in Supplementary Fig. S1a. Furthermore, the shape and size of the tips used in the AFM measurements were determined via three-dimensional imaging using Bruker's tip qualification procedure on a rough Ti sample$^{36}$ (QNM kit) -- see Supplementary Fig. S1b. This allowed us to find the tip radius, $R_{tip}$, for each value of $\delta$ and calculate the contact area, $S$, using a spherical tip approximation:
\begin{align}
S = \pi (\delta ^2 + R_{tip}^2) \nonumber
\end{align}
Different probes with similar $k = 125$ and $127$ N/m (found using Sader's method$^{37}$) were employed in measurements of several bubbles, and the results were independent of the probe within our experimental error. We also verified that the shape and size of the AFM tip did not change during the indentation experiments by repeatedly checking it before and after measurements.

The FDCs were obtained at the centers of graphene and MoS$_2$ bubbles. To find the indentation depth, $\delta$, for each value of the force, $F$, we used reference curves obtained on a non-deforming substrate (sapphire). As shown in Supplementary Fig. S2, $\delta$ was calculated as the tip displacement with respect to the sapphire substrate. Repeated loading/unloading cycles on the same bubble showed high reproducibility, reversible behavior and no signatures of fatigue (Supplementary Fig. S2).  For all the pressure measurements shown in Fig. \ref{fig:pressure} we used a loading/unloading rate of 20 nm/s.

\section*{Acknowledgments.}
We acknowledge funding from the European Commission under the Graphene Flagship, contract CNECTICT- 604391. F. G. is partially funded by ERC, grant 290846, and MINECO (Spain), grant FIS2014-57432.

\section*{References}

1.	Geim, A. K., Grigorieva, I. V. Van der Waals heterostructures. \textit{Nature} \textbf{499,} 419--25 (2013).

\vspace{1mm}

2.	Haigh, S. J. et al. Cross-sectional imaging of individual layers and buried interfaces of graphene-based heterostructures and superlattices. \textit{Nat. Mater.} \textbf{11,} 764--767 (2012).

\vspace{1mm}

3.	Kretinin, A. V. et al. Electronic properties of graphene encapsulated with different two-dimensional atomic crystals. \textit{Nano Lett.} \textbf{14,} 3270--3276 (2014).

\vspace{1mm}

4.	Nair, R. R., Wu, H. A., Jayaram, P. N., Grigorieva, I. V., Geim,  A. K. Unimpeded permeation of water through helium-leak-tight graphene-based membranes. \textit{Science} \textbf{335}, 442--444 (2012).

\vspace{1mm}

5.	Holt, J., Park, H., Wang, Y. Fast mass transport through sub-2-nanometer carbon nanotubes. \textit{Science} \textbf{312}, 1034--1038 (2006).

\vspace{1mm}

6.	Kolesnikov, A. I. et al. Anomalously soft dynamics of water in a nanotube: A revelation of nanoscale confinement. \textit{Phys. Rev. Lett.} \textbf{93}, 035503 (2004).

\vspace{1mm}

7.	Yuk, J. M. et al. High-resolution EM of colloidal nanocrystal growth using graphene liquid cells. \textit{Science} \textbf{336}, 61--64 (2012).

\vspace{1mm}

8.	Park, J. et al. Direct observation of wet biological samples by graphene liquid cell transmission electron microscopy. \textit{Nano Lett.} \textbf{15}, 4737--4744 (2015).

\vspace{1mm}

9.	Wojcik, M., Hauser, M., Li, W., Moon, S., Xu, K. Graphene-enabled electron microscopy and correlated super-resolution microscopy of wet cells. \textit{Nat. Commun.} \textbf{6}, 7384 (2015).

\vspace{1mm}

10.	Xu, K., Cao, P., Heath, J. R. Graphene visualizes the first water adlayers on mica at ambient conditions. \textit{Science} \textbf{329}, 1188--1191 (2010).

\vspace{1mm}

11.	Algara-Siller, G. et al. Square ice in graphene nanocapillaries. \textit{Nature} \textbf{519}, 443--445 (2015).

\vspace{1mm}

12.	Lim, C. H. Y. X. et al. A hydrothermal anvil made of graphene nanobubbles on diamond. \textit{Nat. Commun.} \textbf{4}, 1556 (2013).

\vspace{1mm}

13.	Levy, N. et al. Strain-induced pseudo-magnetic fields greater than 300 tesla in graphene nanobubbles. \textit{Science} \textbf{329}, 544--547 (2010).

\vspace{1mm}

14.	Guinea, F., Katsnelson, M. I., Geim, A. K. Energy gaps and a zero-field quantum Hall effect in graphene by strain engineering. \textit{Nat. Phys.} \textbf{6}, 30--33 (2010).

\vspace{1mm}

15.	Bertolazzi, S., Brivio, J., Kis, A. Stretching and breaking of ultrathin MoS$_{2}$. \textit{ACS Nano} \textbf{5}, 9703--9709 (2011).

\vspace{1mm}

16.	Peng, Q., De, S. Outstanding mechanical properties of monolayer MoS$_{2}$ and its application in elastic energy storage. \textit{Phys. Chem. Chem. Phys.} \textbf{15}, 19427--37 (2013).

\vspace{1mm}

17.	Landau L. D. and Lifschitz E. M. \textit{Theory of Elasticity.} Pergamon Press, New York, 1970.

\vspace{1mm}

18. Nelson, D. R., Piran,  T. Weinberg, S.  eds., \textit{Statistical properties of membranes and surfaces.} World Scientific, Singapore, 2004.

\vspace{1mm}

19. King, H., Schroll, R. D., Davidovitch,  B., Menon,  N. Elastic sheet on a liquid drop reveals wrinkling and crumpling as distinct symmetry-breaking instabilities. \textit{PNAS} \textbf{109}, 9716--9720 (2012).

\vspace{1mm}

20.	Koenig, S. P. Boddeti, N. G. Dunn, M. L. Bunch, J. S. Ultrastrong adhesion of graphene membranes. \textit{Nat. Nanotechnol.} \textbf{6}, 543--546 (2011).

\vspace{1mm}

21.	Bj\"orkman, T., Gulans,  A., Krasheninnikov,  A. V., Nieminen, R. M. Van der Waals bonding in layered compounds from advanced density-functional first-principles calculations. \textit{Phys. Rev. Lett.} \textbf{108}, 1--5 (2012).

\vspace{1mm}

22.	Sachs, B., Wehling, T. O., Katsnelson, M. I., Lichtenstein, A. I. Adhesion and electronic structure of graphene on hexagonal boron nitride substrates. \textit{Phys. Rev. B} \textbf{84}, 195414 (2011).

\vspace{1mm}

23.	Peng, Q. Ji, W. De, S. Mechanical properties of the hexagonal boron nitride monolayer: Ab initio study. \textit{Comput. Mater. Sci.} \textbf{56}, 11--17 (2012).

\vspace{1mm}

24. Castellanos-Gomez, A., Poot, M., Steele, G. A., van der Zant, H. S., Agra\"it, N., Rubio-Bollinger, G. Elastic properties of freely suspended MoS$_{2}$2 nanosheets. \textit{Adv. Mater.} \textbf{24} 772--775 (2012).

\vspace{1mm}

25. Nicholl, R. J. et al. The effect of intrinsic crumpling on the mechanics of free-standing graphene. \textit{Nature Commun.} \textbf{6}, 8789 (2015).

\vspace{1mm}

26.	Eckmann, A. et al. Raman fingerprint of aligned graphene/h-BN superlattices. \textit{Nano Lett.} \textbf{13}, 5242--5246 (2013).

\vspace{1mm}

27.	Gibertini, M., Tomadin, A., Guinea, F., Katsnelson, M. I., Polini, M. Electron-hole puddles in the absence of charged impurities. \textit{Phys. Rev. B} \textbf{85}, 201405 (2012).

\vspace{1mm}

28. Roos, W. H.,  Bruinsma, R., Wuite,  G. J. L. Physical virology. \textit{Nat. Phys.} \textbf{6}, 733--743 (2010).

\vspace{1mm}

29. G. J. Orssengo and D. C. Pye, Determination of the true intraocular pressure and modulus of elasticity of the human cornea in vivo. \textit{Bull. Math. Biol.} \textbf{61}, 551--572 (1999).

\vspace{1mm}

30. Lee, C., Wei, X. D., Kysar, J. W.,  Hone, J. Measurement of the elastic properties and intrinsic strength of monolayer graphene. \textit{Science} \textbf{321}, 385-388 (2008).

\vspace{1mm}

31. L\'opez-Pol\'in, G., Jaafar, M., Guinea, F., Rold\'an, R., G\'omez-Navarro, C., G\'omez-Herrero, J. Strain dependent elastic modulus of graphene. Preprint at arXiv:1504.05521 (2015).

\vspace{1mm}

32. Los, J. H., Fasolino, A., Katsnelson, M. I.  Scaling behavior and strain dependence of in-plane elastic properties of graphene. \textit{Phys. Rev. Lett.} \textbf{116}, 015901 (2016).

\vspace{1mm}

33. Butt, H. J., Cappella, B. Kappl, M. Force measurements with the atomic force microscope: Technique, interpretation and applications. \textit{Surf. Sci. Rep.} \textbf{59}, 1--152 (2005).

34.	Dokukin, M. E. Sokolov, I. On the measurements of rigidity modulus of soft materials in nanoindentation experiments at small depth. \textit{Macromolecules} \textbf{45}, 4277--4288 (2012).

\vspace{1mm}

35. Sokolov, I. et al. AFM study of forces between silica, silicon nitride and polyurethane pads. \textit{J. Colloid Interface Sci.} \textbf{300}, 475--481 (2006).

\vspace{1mm}

36. Hua, Y. PeakForce-QNM advanced applications training 2014. \url{http://mmrc.caltech.edu/AFM Dimension Icon/Bruker Training}.

\vspace{1mm}

37.	Sader, J. E., Larson, I., Mulvaney, P. White, L. R. Method for the calibration of atomic force microscope cantilevers. \textit{Rev. Sci. Instrum.} \textbf{66}, 3789--3798 (1995).

\end{document}

%\preprint{}

\title{Graphene bubbles on a substrate: Universal shape and van der Waals pressure. Supplementary Information.}
\author{E. Khestanova}
\affiliation{School of Physics and Astronomy, University of Manchester, Oxford Road, Manchester M13 9PL, UK}
\author{F. Guinea$^*$}
\affiliation{School of Physics and Astronomy, University of Manchester, Oxford Road, Manchester M13 9PL, UK}
\affiliation{IMDEA Nanociencia, Faraday, 9, Cantoblanco, 28049, Madrid, Spain}
\author{L. Fumagalli}
\affiliation{School of Physics and Astronomy, University of Manchester, Oxford Road, Manchester M13 9PL, UK}
\author{A. K. Geim}
\affiliation{School of Physics and Astronomy, University of Manchester, Oxford Road, Manchester M13 9PL, UK}
\author{I. V. Grigorieva$^*$}
\affiliation{School of Physics and Astronomy, University of Manchester, Oxford Road, Manchester M13 9PL, UK}

\maketitle

\begin{figure}[h]
	\includegraphics[width=0.35\textwidth]{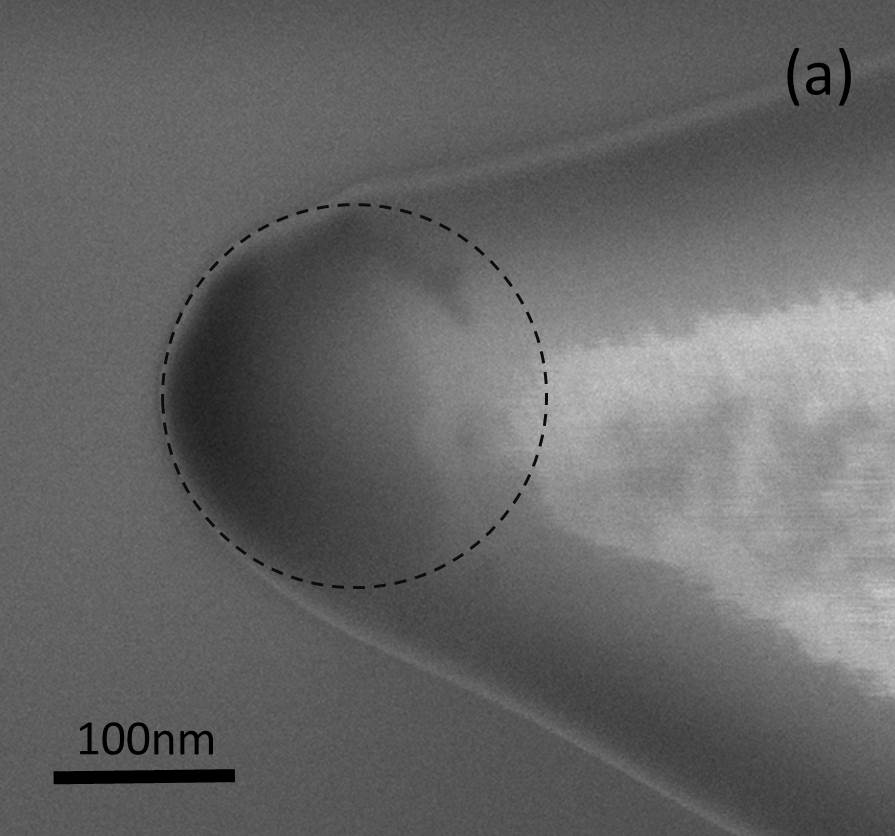}\hspace{5mm}
	\includegraphics[width=0.52\textwidth]{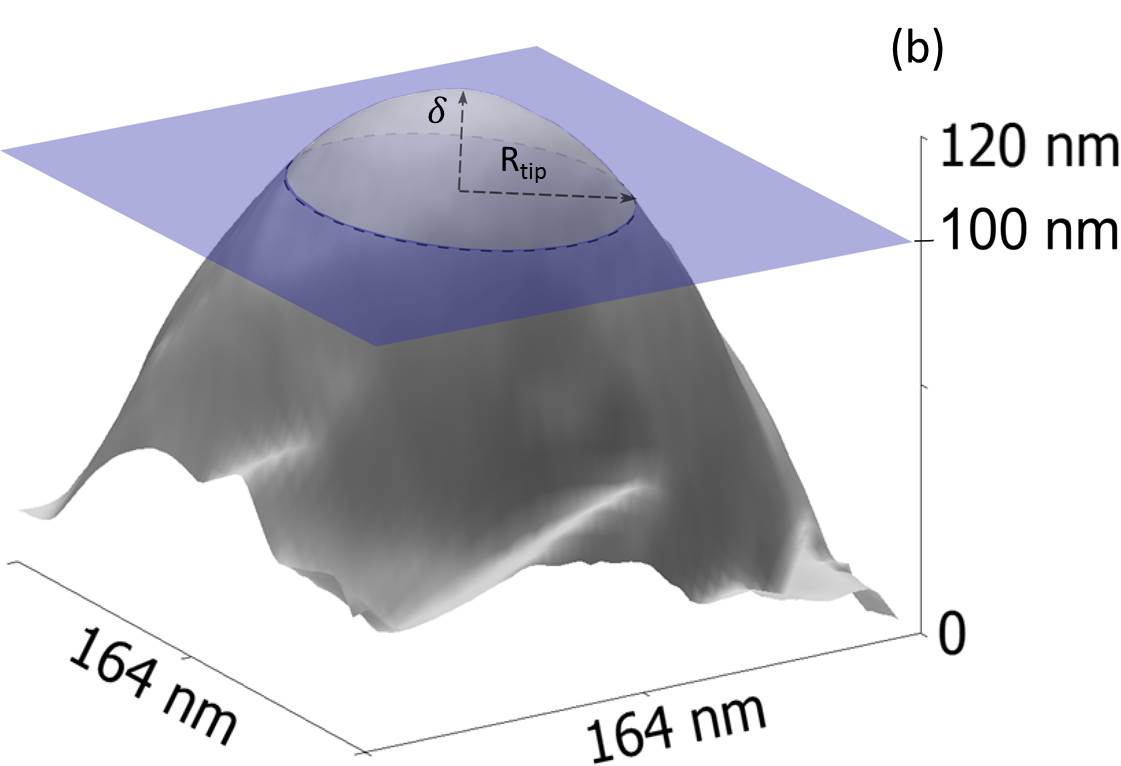}\vspace{10mm}
		{\textbf{Supplementary Fig. S1}. \textbf{AFM tips used in indentation experiments.} \textbf{(a)} Scanning electron microscope (SEM) image of one of the AFM tips. The dashed circle emphasises its spherical shape. \textbf{(b)} Three-dimensional image of another AFM tip obtained using the Bruker calibration procedure (see Methods). The plane section indicates the maximum range of indentation depths corresponding to applicability of our numerical fitting procedure.}
\end{figure}

\begin{figure}[t]
	\includegraphics[width=0.6\textwidth]{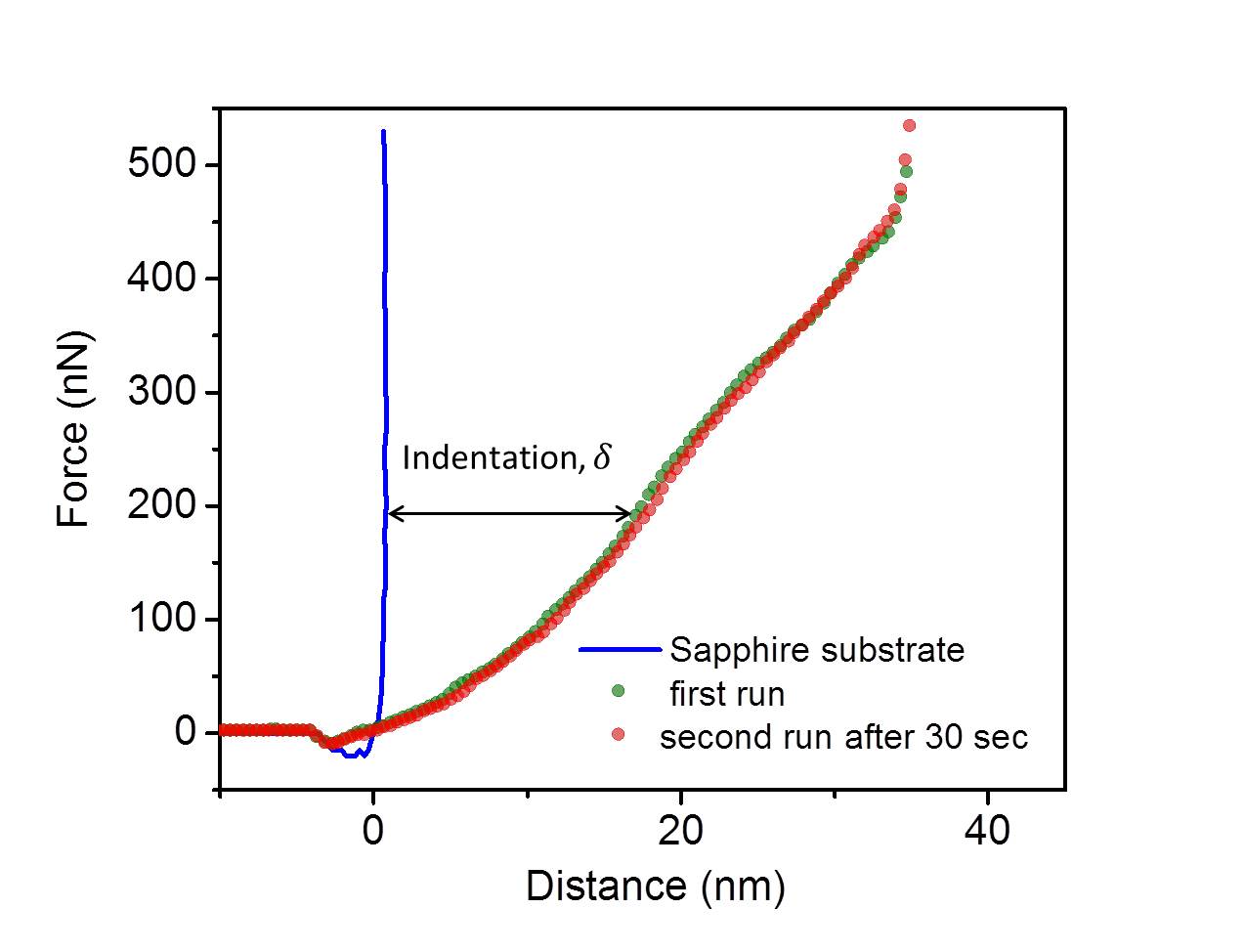}\vspace{5mm}
	
{\textbf{Supplementary Fig. S2.} \textbf{Reproducibility of force-distance curves}. Two force-distance curves taken on the same graphene bubble ($R=$317 nm). The blue line shows the force-distance curve taken with the same AFM tip on a sapphire substrate.}
\end{figure}

\clearpage

\section*{Supplementary Note 1.}
\label{sec:elastic_energy_relaxation}
\subsection*{Elastic Energy.}
The elastic energy of one bubble, defined by a profile $h ( \vec{\bf r} )$ is
\begin{align}
E_{el} = \frac{\mu ( \lambda + \mu )}{2 ( \lambda + 2 \mu )} \frac{1}{4 \pi^2} \int d^2 \vec{\bf q} \left| \sum_{i,j} \left( 1 - \frac{q_i q_j}{| \vec{\bf q} |^2} \right) f_{i,j} ( \vec{\bf q} ) \right|^2
\label{erelax}
\end{align}
where $i,j = x,y$ and
\begin{align}
f_{ij} ( \vec{\bf q} ) &= \int d^2 \vec{\bf r} e^{i \vec{\bf q} \vec{\bf r}} \partial_i h ( \vec{\bf r} ) \partial_j h ( \vec{\bf r} )
\end{align}
For an isotropic bubble, $h ( \vec{r} ) = h ( r )$, where $r = | \vec{\bf r} |$, equation (\ref{erelax}) becomes
\begin{align}
E_{el} &= \frac{\pi}{8} \frac{\mu ( \lambda + \mu )}{\lambda + 2 \mu} \times  \int_0^\infty dq q \left| \int_0^\infty dr r \left[ J_0 ( q r ) - J_2 ( q r ) \right] \left( \partial_r h \right)^2 \right|^2
\label{erelax2}
\end{align}
where $J_0 ( u )$ and $J_2 ( u )$ are Bessel functions.

In the presence of an isotropic strain, $\epsilon_{xx} = \epsilon_{yy} = \epsilon / 2$, the bubble acquires an additional energy,
\begin{align}
E_{str} &=  \frac{\epsilon \mu ( \lambda + \mu )}{\lambda + 2 \mu }  \lim_{\vec{\bf q} \rightarrow 0} \sum_{i,j} \left( 1 - \frac{q_i q_j}{| \vec{\bf q} |^2} \right) f_{i,j} ( \vec{\bf q} )
\end{align}
where we have used the relation $\epsilon ( \vec{\bf q} ) = 4 \pi^2 \epsilon \delta^2 ( \vec{\bf q} )$. For an isotropic height profile, $h ( r )$, equation (\ref{erelax}) reduces to
\begin{align}
E_{str} &= \frac{\pi \epsilon \mu ( \lambda + \mu )}{2 ( \lambda + 2 \mu )} \int_0^\infty dr r \left[ \partial_r h ( r ) \right]^2
\end{align}

Alternatively, for an isotropic height profile, $h ( r )$, one can write the elastic energy as
\begin{align}
E_{el} &= 2 \pi \frac{\lambda}{2}  \int_0^\infty dr r \left[ \partial_r u_r + \frac{u_r}{r} + \frac{( \partial_r h )^2}{2} \right]^2 + 2 \pi \mu  \int_0^\infty dr r \left[ \left( \partial_r u_r + \frac{( \partial_r h )^2}{2} \right)^2 + \frac{u_r^2}{r^2} \right]
\label{e_elas}
\end{align}
The displacement $u_r$ can be obtained from
\begin{align}
- &( \lambda + 2 \mu ) \left( \partial_r^2 u_r + \frac{\partial_r u_r}{r} - \frac{u_r}{r^2} \right) =  ( \lambda + 2 \mu ) \left( \partial_r^2 h \right) \left( \partial_r h \right) + \frac{\mu}{r} \left( \partial_r h \right)^2
\label{ur}
\end{align}

\section*{Supplementary Note 2.}
\subsection*{Influence of tensile strains.}
Tensile strains, $\epsilon > 0$, modify the aspect ratio of the bubbles, as shown in equation (7) in the main text. The changes significant  when the term proportional to $\epsilon$ becomes a significant fraction of the total elastic energy. By comparing these contributions in equation (7), we find a threshold strain, $\epsilon_{th}$
\begin{align}
\epsilon_{th} &\approx \frac{c_1}{c_2} \left( \frac{h_{max}}{R} \right)^2 = \frac{c_1}{c_2} \left( \frac{\pi \gamma}{5 Y} \right)^{1/2}
\label{eth}
\end{align}
Using $\gamma \approx 0.015$ eV, and the Young modulii for graphene and MoS$_2$ we find, for graphene, $\epsilon_{th} \approx 1.4 \%$, and for MoS$_2$, $\epsilon_{th} \approx 1.9 \%$. For larger tensile strains,the aspect ratio tends to
\begin{align}
\frac{h_{max}}{R} \approx \sqrt{\frac{\pi \gamma}{2 c_2 \epsilon Y}}
\end{align}
The bubbles become shorter in the presence of a tensile strain. The dependence of the pressure on the height of the bubble, equation (18) in the main text is not significantly modified, and we find
\begin{align}
P &\approx \frac{\pi \gamma}{c_V h_{max}}
\end{align}

\subsection*{Influence of compressive strains.}
In the presence of compressive strains, $\epsilon < 0$, the aspect ratio $h_{max} / R$, increases. A sufficiently high compressive strain will overcome the vdW interaction, and the graphene layer will delaminate, forming a bubble even in the absence of trapped material, or can even completely detach from the substrate. If we do not consider the internal energy of the material inside the bubble, a bubble formed spontaneously with radius $R$ and height $h_{max}$ has an energy,

\begin{align}
E_{tot} &=  c_1 Y \frac{h_{max}^4}{R^2} - c_2 Y | \epsilon | h_{max}^2  + \pi \gamma R^2
\label{elayer2}
\end{align}
Minimizing this expression with respect to $h_{max}$, we find
\begin{align}
\frac{h_{max}^2}{R^2} &= \frac{c_2 | \epsilon |}{2 c_1}
\end{align}
The total energy of the bubble, as function of $R$, is
\begin{align}
E_{tot} &= - \frac{Y c_2^2 | \epsilon|^2}{4 c_1} R^2 +\pi \gamma R^2.
\end{align}
The bubble will be stable and grow towards $R \rightarrow \infty$ if the compressive strain is such that
\begin{align}
\frac{Y c_2^2 | \epsilon|^2}{4 c_1} > \pi \gamma
\end{align}
The compressive strain required for the bubble to be unstable is
\begin{align}
| \epsilon | &\ge \sqrt{\frac{4 c_1 \pi \gamma}{Y c_2^2}}
\end{align}
These strains are $6.2 \%$ and $8.7 \%$ for graphene and MoS$_2$, respectively.

On the other hand, for sufficiently small strains, we can expand the solution of equation (8) in the main text, and we obtain
\begin{align}
\frac{h_{max}}{R} &\approx \left( \frac{c_4 \gamma}{5 c_1 Y} \right)^{1/4} - \frac{c_2  \epsilon }{10 c_1} \sqrt{\frac{5 c_1 Y}{\pi \gamma}},
\label{compressive}
\end{align}
that is, for $\epsilon < 0$ (compressive strain)  $h_{max} / R$ will be greater than in the unstrained situation.

\subsection*{Interaction between bubbles.}
Bubbles can only interact when the in plane displacements relax to the height profile. We can extend eqs.(\ref{erelax}) and (\ref{erelax2}) to the case of two non overlapping profiles, $h_1 ( \vec{\bf r} )$ and $h_2 ( \vec{\bf r} )$. Leaving out the self energy of each bubble, we obtain

\begin{align}
E_{int} &= \frac{\mu ( \lambda + \mu )}{2 ( \lambda + 2 \mu )} \frac{1}{4 \pi^2} \int d^2 \vec{\bf q} e^{i \vec{\bf q} \vec{\bf R}}  \times  \left[ \sum_{i,j} \left( 1 - \frac{q_i q_j}{| \vec{\bf q} |^2} \right) f_{i,j}^1 ( \vec{\bf q} ) \right] \left[ \sum_{i,j} \left( 1 - \frac{q_i q_j}{| \vec{\bf q} |^2} \right) f_{i,j}^2 ( \vec{\bf q} )  \right]
\label{eint}
\end{align}
where $\vec{\bf R}$ is the vector connecting the centers of the two bubbles, and $f_{i,j}^k ( \vec{\bf q} )$ is, as before, the Fourier transform of $\partial_i h_k ( \vec{\bf r} ) \partial_j  h_k ( \vec{\bf r} )$.

\begin{align}
E_{int} ( R ) &= \frac{\pi}{8} \frac{\mu ( \lambda + \mu )}{\lambda + 2 \mu}  \int_0^\infty dq q J_0 ( q R ) \times \left\{ \int_0^\infty dr r \left[ J_0 ( q r ) - J_2 ( q r )
\right] \left( \partial_r h_1 \right)^2  \right. \times \left.   \int_0^\infty dr r \left[ J_0 ( q r ) - J_2 ( q r )
\right] \left( \partial_r h_2 \right)^2 \right\}
\end{align}
where $R = | \vec{\bf R} |$. A long distances $R  \gg \ell$, we obtain
\begin{align}
E_{int} ( R ) &\approx  \frac{\alpha \pi}{8} \frac{\mu ( \lambda + \mu )}{( \lambda + 2 \mu ) R^2 } \int_0^\infty dr r \left( \partial_r h_1 \right)^2  \int_0^\infty dr r \left( \partial_r h_2 \right)^2
\end{align}
where
\begin{align}
\alpha &= \lim_{q \rightarrow 0} r^2 \int_0^\infty dq q J_0 ( q r ) \approx -1.11 \times 10^{-6}
\end{align}
The interaction between bubbles is attractive, favouring their merger into bigger bubbles, as observed experimentally.